\documentclass[lettersize,journal]{IEEEtran}
\usepackage[font=small,justification=centering]{caption}

\usepackage[caption=false,font=footnotesize,labelfont=sf,textfont=sf]{subfig}	% font = \tiny, \scriptsize, \footnotesize, \small, \normalsize, \large, \Large, \LARGE, \huge, \Huge
\usepackage{array}
\usepackage{graphicx}
\usepackage{algorithmic}
\usepackage{algorithm}
\usepackage{bm}
\usepackage{textcomp}
\usepackage{cite}
\usepackage{amsmath, amsfonts,amssymb, amsthm}
\usepackage{wasysym}
\usepackage{booktabs}
\usepackage{paralist}
\usepackage{balance}
\usepackage{multicol}
\usepackage{mathrsfs}
\usepackage{epsfig}
\usepackage{epstopdf}
\usepackage{upgreek}
\usepackage{color}
\usepackage{enumerate}
\usepackage{enumitem}
\usepackage{url}
\usepackage{verbatim}
\usepackage{graphicx}
\usepackage{mathrsfs}
\usepackage{tabularx}
\def\BibTeX{{\rm B\kern-.05em{\sc i\kern-.025em b}\kern-.08em
		T\kern-.1667em\lower.7ex\hbox{E}\kern-.125emX}}
\graphicspath{ {fig/} }

\setenumerate{leftmargin = 0em, itemindent = 15pt, listparindent = 1em}

\begin{document}

\title{VNF Migration with Fast Defragmentation: A GAT-Based Deep Learning Method}
\author{Fangyu~Zhang,~\IEEEmembership{Graduate Student Member,~IEEE},
		Yuang~Chen,~\IEEEmembership{Graduate Student Member,~IEEE},
		Hancheng~Lu,~\IEEEmembership{Senior Member,~IEEE},
		Chengdi~Lu
\IEEEcompsocitemizethanks{
	\IEEEcompsocthanksitem Fangyu Zhang, Yuang Chen, Hancheng Lu, and Chengdi Lu are with CAS Key Laboratory of Wireless-Optical Communications, University of Science and Technology of China, Hefei 230027, China (email: fv215b@mail.ustc.edu.cn; yuangchen21@mail.ustc.edu.cn; hclu@ustc.edu.cn; lcd1999@mail.ustc.edu.cn).
}
}
\maketitle
\begin{abstract}
Network function virtualization (NFV) enhances service flexibility by decoupling network functions from dedicated hardware. To handle time-varying traffic in NFV network, virtualized network function (VNF) migration has been involved to dynamically adjust resource allocation. However, as network functions diversify, different resource types may be underutilized due to bottlenecks, which can be described as multidimensional resource fragmentation. To address this issue, we firstly define a metric to quantify resource fragmentation in NFV networks. Then, we propose a multi-hop graph attention network (MHGAT) model to effectively extract resource features from tailored network layers, which captures the overall network state and produces high-quality strategies rapidly. Building on this, we develop an MHGAT method to implement fast defragmentation and optimize VNF migration. Simulations demonstrate that by fast defragmentation, the MHGAT method improves the acceptance ratio by an average of 12.8\%, reduces the overload ratio by an average of 30.6\%, and lowers migration loss by an average of 43.3\% compared to the state-of-art benchmark.
\end{abstract}

\IEEEdisplaynontitleabstractindextext

\IEEEpeerreviewmaketitle
\vspace{-3mm}
\section{Introduction}
Network function virtualization (NFV) is an emergent network architecture advocated by European Telecommunications Standards Institute (ETSI), aimed at overcoming the limitations of traditional network devices through the use of standard virtualization technologies \cite{yi2020design}. Specifically, by decoupling network functions, such as firewalls, network address translation, etc., from dedicated hardware in the form of virtual machines or containers, NFV creates the potential for network service providers to increase the flexibility of deploying new services in response to expanding and diversifying customer demands. Driven by the desire to reduce the cost of network deployment and increase the speed of service deployment, NFV is receiving attention from headline telcos such as Huawei, Ericsson, and Nokia, and is expected to play a significant role in new network architectures such as 5G network slicing, radio access network (RAN), and 6G \cite{etsi2023evolving}.

\begin{figure}[!t]
	\centering
	\includegraphics[width=3.5in]{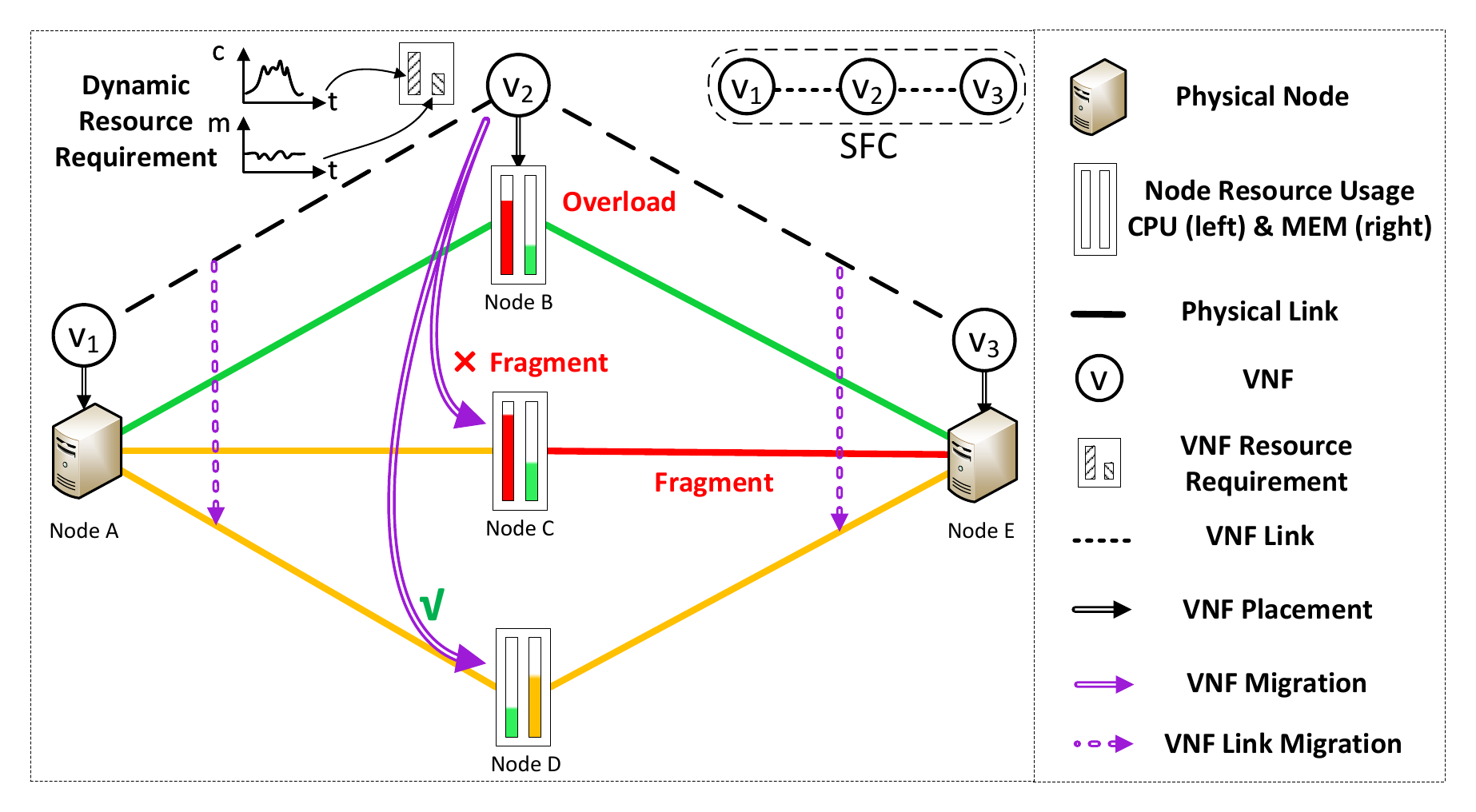}
	\caption{VNF migration in NFV network.}
	\label{fig_mig}
\end{figure}

Network functions virtualized using NFV technology, i.e., virtualized network functions (VNFs), are usually composed into service function chains (SFCs) to provide a complete service flow according to a specific graph structure, i.e., the VNF-forwarding graph (VNF-FG). During SFC operation, due to the time-varying nature of user traffic, the allocation of infrastructure resources typically needs to be adjusted to achieve goals such as load balancing, energy saving, and fault tolerance \cite{afra2024cost}. One solution that is considered to be potential to achieve these goals is VNF migration, i.e., moving a VNF instance from one physical host to another. Fig. \ref{fig_mig} illustrates VNF migration in an NFV network. A large amount of work has been done to investigate how VNF migration can be performed to improve quality of service (QoS) for users or to reduce capital expenditure (CAPEX) and operating expenditure (OPEX) for service providers \cite{zhang2021online, wen2022failure, shang2022online}. However, there are still serious issues in the work of reducing physical machine overload or load balancing achieved by VNF migration.

A pressing issue is the resource fragmentation problem. The resource fragmentation problem refers to the inability to utilize one type of abundant resource in an infrastructure network due to the shortage of another resource. These resources, encompassing different categories associated with both links and nodes, are referred to as multidimensional resources, such as bandwidth \cite{wang2022multi}, CPU \cite{pham2022optimizing}, memory \cite{ali2023multi}, and even graphics processing units (GPUs) \cite{zeng2023enable}. Several studies have been conducted to demonstrate that resource fragmentation leads to load imbalance in an implicit manner, which reduces the resource utilization of the network and the acceptance rate of new services \cite{lu2022resource, bao2022sdfa, lar2023impact, ma2023spatio}. However, in the VNF migration problem, the network's multidimensional resource utilization fluctuates all the time due to the time-varying VNF traffic and the varying amount of multidimensional resources consumed by individual VNF instances, which makes it challenging to obtain a defragmentation-capable VNF migration strategy in a short period of time.

To quickly address the issue of resource fragmentation, efficient VNF migration strategies that are aware of fragmentation are required. However, the methods for perceiving resource fragmentation and efficiently migrating VNFs have not been thoroughly researched. On the one hand, although some studies have used the weighted sum of the variance of multidimensional resources as a metric of resource fragmentation, this metric struggles to effectively reflect the level of fragmentation \cite{tseng2018dynamic, chen2024fault,liu2023deep}. For example, when different types of resources have similar variances but different utilization rates, the resource with higher utilization becomes a bottleneck, leading to fragmentation. That is to say, the weighted sum of variances can only represent variance information. On the other hand, while traditional heuristic migration algorithms can solve the NP-hard VNF migration problem in polynomial time, they face a trade-off between the number of iterations and performance \cite{li2021joint, ali2023multi,liao2024collab}. Emerging deep reinforcement learning (DRL) techniques eliminate this barrier by pre-training deep neural networks, allowing for fast generation of optimal migration strategies after training \cite{wang2024cost, pham2022optimizing, qu2022prio, he2024load, liu2023deep, afra2023rein}. However, the raw neural network architecture design and the interactive nature of DRL limit both the training speed and operational efficiency of the neural networks involved \cite{pham2022optimizing,afra2023rein,wang2024cost}. Therefore, to achieve fast defragmentation through VNF migration, new resource fragmentation-aware metric is needed, along with improvements in the speed of VNF migration decision-making.

In this paper, we define a new metric to quantitatively measure multidimensional resource fragmentation and propose a novel deep learning method to rapidly solve the VNF migration problem. First, we build a fragmentation model to represent the quantified multidimensional resource fragmentation of network nodes, which can perceive the multidimensional resource information of neighbors at different distances. Then, we propose a metric named maximum weighted fragmentation level to measure the fragmentation level of the network and formalize the VNF migration problem with the objective of reducing this metric and migration loss. To solve this problem, we propose a novel neural network model, i.e. multi-hop graph attention network (MHGAT), to quickly obtain the new defined metrics with different migration strategies. Based on the MHGAT model, an MHGAT method is proposed to solve the VNF migration problem efficiently with fast defragmentation. The main contributions of this paper are summarized as follows.
\begin{itemize}
	\item We propose a node fragmentation model to characterize the multidimensional resource information within the multi-hop distance of that node. Combining the model with the overall resource conditions of the network, we derive the maximum weighted fragmentation level as a new metric to quantify the maximum resource fragmentation in the network. Simulation results validate the correlation between the new defined metric and network overload ratio.
	
	\item Based on the defined metric, we construct a novel MHGAT model to efficiently extract the multidimensional resource features of the network and rapidly output the maximum weighted fragmentation level of the network with different migration strategies. The results of ablation experiments show that all modules of the proposed MHGAT model improve the neural network performance.
	
	\item Based on the MHGAT model, we propose an MHGAT method to solve the VNF migration problem with fast defragmentation. It controls the migration loss through VNF selection and uses the MHGAT model to obtain a defragmentation policy. Simulation results demonstrate that the MHGAT method outperforms the state-of-the-art benchmark in terms of fast defragmentation.
\end{itemize}
The remainder of the paper is organized as follows. Section II discusses the work related to VNF migration and defragmentation. Section III gives the system model modeling and the proposed fragmentation model. In Section IV we propose the new metric and formalize the VNF migration problem with the objective of minimizing the new metric and migration loss. Section V details our proposed deep learning approach. Simulation results are discussed in Section VI. Finally, the paper is concluded in Section VII.
\vspace{-2.5mm}
\section{Related Work}
In this section, we present existing related work in VNF migration and defragmentation.
\vspace{-3.5mm}
\subsection{VNF Migration}

The objectives of the VNF migration work mainly include reducing costs, energy consumption, and failures \cite{tang2019virtual, hu2023migration, zhu2023avail, zhai2023migration}. Tang \textit{et al}. \cite{tang2019virtual} proposed a deep belief network (DBN) based prediction method to predict the changes in network load and a heuristic algorithm to perform VNF migration to minimize the cost of resource usage. Hu \textit{et al}. \cite{hu2023migration} proposed a VNF migration method based on long and short-term memory (LSTM) networks to simultaneously reduce energy consumption and migration loss. Zhu \textit{et al}. \cite{zhu2023avail} proposed an SFC availability model to consider service availability assessment after VNF migration and validated the proposed availability model through VNF migration. Zhai \textit{et al}. \cite{zhai2023migration} proposed a fault prediction method based on improved LSTM to predict the failure of network nodes and a meta-heuristic to migrate VNFs in advance based on the prediction results.

Although most of the research has centered around the aforementioned objectives, new network architectures or scenarios, such as multi-access edge computing (MEC) and heterogeneous networks, pose new problems \cite{emu2021ensemble, bai2023uav, li2023qos}. Emu \textit{et al}. \cite{emu2021ensemble} consider the use of VNFs on cloudlets located at the network edge and propose a deep learning based VNF deployment approach to reduce communication costs. Bai \textit{et al}. \cite{bai2023uav} considered the vulnerability of deploying VNFs in unmanned aerial vehicle assisted multi-access edge computing (UMEC) network, and proposed a quantitative modeling approach based on semi-Markov processes to explore the availability resilience of SFCs in UMEC networks. Li \textit{et al}. \cite{li2023qos} considered the VNF migration problem in heterogeneous networks with programmable data plane switches applied and proposed a layered auxiliary graph-based approach to improve user QoS.

Since much of the work on VNF migration needs to optimize multiple objectives and VNF migration requires real-time performance, the algorithms for solving the problem mainly pursue efficiency rather than optimality. The mainstream solution methods are heuristic algorithms \cite{li2021joint, ali2023multi,liao2024collab} and DRL methods \cite{wang2024cost, he2024load, chen2024fault}, both of which are characterized by fast solutions. For heuristic algorithms, Kyoomars \textit{et al}. \cite{ali2023multi} proposed a multi-objective genetic algorithm, a meta-heuristic algorithm, to reduce operational costs through VNF migration. Li \textit{et al}. \cite{li2021joint} proposed a meta-heuristic algorithm and a heuristic algorithm with less computational overhead to minimize the service delay affected by VNF migration. Liao \textit{et al}. \cite{liao2024collab} proposed a heuristic algorithm to obtain a VNF migration policy with the objective of minimizing decision time and system delay in a multi-access edge computing (MEC) network. For DRL methods, Wang \textit{et al}. \cite{wang2024cost} proposed a DRL-based SFC migration method to reduce migration loss and maintain a balanced distribution of resources. He \textit{et al}. \cite{he2024load} proposed a DRL method based on graph attention network (GAT) \cite{vel2018graph} to improve the service acceptance rate and load fairness of satellite networks. Chen \textit{et al}. \cite{chen2024fault} proposed a fault-tolerant oriented multi-agent DRL method to reduce the probability of centralized controller failures by decentralizing the training task from the control plane to the data plane. Although DRL-based migration methods are well known and rapidly evolving for their environmental adaptability and learning intelligence, most of the work still uses only fully connected network as their neural network model, which lags behind the neural network models developed in the field of deep learning, such as graph convolutional network (GCN) \cite{thomas2017semi}, GAT \cite{he2023leveraging, wu2024graph, he2024load}, and graph isomorphism network (GIN) \cite{xu2019how}. Halting at fully connected neural networks leads to limitations in the training speed and network accuracy of DRL methods.
\vspace{-3.5mm}
\subsection{Defragmentation}
The phenomenon of resource fragmentation has been recognized in different domains as seriously affecting the balanced distribution of system load \cite{weng2023beware, khaleel2023efficient, lar2023impact, bao2022sdfa, ma2023spatio}. Bao \textit{et al}. \cite{bao2022sdfa} considered path fragmentation in an elastic optical network infrastructure and proposed a resource allocation scheme to avoid resource fragmentation. Larsson \textit{et al}. \cite{lar2023impact} proposed directed Pod evictions to reduce node resource fragmentation for containers in a Kubernetes managed environment. Ma \textit{et al}. \cite{ma2023spatio} considered the phenomenon of temporal-spatial resource fragmentation in computing power networks (CPNs) and proposed a GAT-based DRL approach to defragmentation. Weng et al. \cite{weng2023beware} considered the GPU resource fragmentation problem due to GPU sharing technology and proposed a new fragmentation metric for quantifying the degree of GPU fragmentation. Khaleel \cite{khaleel2023efficient} proposed a meta-heuristic algorithm to reduce VM resource fragmentation in Internet of Things (IoT) networks. In addition, our previous work \cite{lu2022resource} considered resource fragmentation in the virtual network embedding problem and quantified it for static uni-dimensional node resource fragmentation and uni-dimensional link resource fragmentation. However, in the context of dynamic and multidimensional resource requirements in VNFs, quantitative metrics for uni-dimensional resource fragmentation are not sufficient to characterize resource fragmentation. This motivates us to investigate quantitative metrics for multidimensional resource fragmentation and method that can quickly compute this metric.

Although current research provides valuable insights into VNF migration, generic load balancing metrics \cite{lu2022resource, bao2022sdfa, lar2023impact, ma2023spatio} and fully connected networks in the DRL methods \cite{pham2022optimizing,afra2023rein,wang2024cost} still limit further research. As the variety of VNFs continues to expand, fragmentation among multidimensional resources becomes increasingly complex, making it more challenging for existing metrics to accurately reflect the load balancing state. Additionally, in DRL-based VNF migration methods, mainstream research still relies on fully connected networks for feature extraction, which hinders efficient graph feature extraction.

\vspace{-3.5mm}
\section{System Model}\label{sys_model}
In this section, we give the system model and the multidimensional resource fragment model. We first introduce the NFV network model and the request model of SFC in Sec. \ref{net_model} and Sec. \ref{req_model}, respectively. For a quick understanding of the network model and the request model, we use Fig. \ref{fig_mig} to show the process of adjusting the resource allocation by VNF migration in an NFV network. In Sec. \ref{frag_model}, we formalize the multidimensional resource fragment model, which can represent node resource and link resource information of arbitrary dimensions as a node vector.

\vspace{-4.5mm}
\subsection{NFV Network Model}\label{net_model}
The NFV network is modeled to describe the infrastructure that provides the physical resources for SFCs. We represent an NFV network by an undirected graph $G=(N,E)$, where $N = \{n_1, n_2, \cdots, n_{|N|}\}$ denotes the set of network nodes and $E = \{e_{12}, e_{23}, \cdots, e_{(|N|-1)|N|}\}$ denotes the set of network links. For each network node $n$, we denote its multidimensional resource capacity by a resource capacity vector $\mathbf{r}^n = [r^n_1, r^n_2, \cdots]$. Note that although this vector supports more dimensions, we only consider the most commonly used two-dimensional resource types in this paper: CPU resources and memory resources. For each network link, a vector $\mathbf{r}^e=[r^e_1, \cdots]$ is used to represent its resource capacity. Similarly, only the most commonly used bandwidth resources are considered in this paper. In addition, we use $d^e$ to denote the fixed propagation delay of each network link.
\vspace{-5mm}
\subsection{SFC Request Model}\label{req_model}
SFC request is modeled to describe the dynamic occupation of physical resources by the requester. We denote the set of SFC requests by $S = \{s_1, s_2, \cdots, s_{|S|}\}$, where $s=\{\omega^s, G^s\}$. For each SFC request $s$, we denote its lifetime by $\omega^s$ and describe the VNF topology of SFC $s$ with a directed acyclic graph (DAG) $G^s=(V^s,L^s)$, where $V^s = \{ v^s_1, v^s_2, \cdots, v^s_{|V^s|}\}$ denotes the set of VNFs and $L^s = \{ l^s_{12}, l^s_{23}, \cdots, l^s_{(|V^s|-1)|V^s|}\}$ denotes the set of VNF links. When the lifetime of an SFC is up, this SFC is removed from the network and the resources it utilized are released.

In a dynamic network environment, the resource occupancy of VNFs and VNF links changes over time. We denote the current moment by $t\!\in\! T$, where $T$ denotes the network operation period. For each VNF $v=\{\mathbf{r}^v(t), d^p_v\}$, $\mathbf{r}^v(t)=[r^v_1(t), r^v_2(t)]$ denotes its resource demand at moment $t$, and $d^p_v$ denotes the processing delay incurred by its task. For each VNF link $l=\{\mathbf{r}^l(t), D_l\}$, $\mathbf{r}^l(t)=[r^l_1(t)]$ denotes its resource demand at moment $t$, and $D_l$ denotes its task deadline.
\vspace{-5mm}
\subsection{Multidimensional Resource Fragment Model}\label{frag_model}
In this subsection, we introduce the multidimensional resource fragment model, which can easily reflect the degree of multidimensional resource balancing of the entire network.

First, to represent the utilization of each type of resource, we denote the set of VNFs on node $n$ and the set of VNF links on link $e$ by $V^n$ and $L^e$, respectively. Then we can represent the resource utilization of type $i$ on node $n$ as follows:
\begin{equation}
	u^n_i(t) = \sum_{v\in V^n}r^v_i(t).
\end{equation}

Similarly, the resource utilization of type $i$ on a network link $e$ can be denoted as
\begin{equation}
	u^e_i(t) = \sum_{l\in L^e}r^l_i(t).
\end{equation}

Then, we define the receptive field $k$ of the fragment model, which represents the range of each node's perception of its neighboring nodes' resources. For example, when $k=2$, the fragment model of node $n$ considers neighboring nodes and paths that are two hops away from node $n$. 

To analyze the resource connectivity of the $k$-hop paths adjacent to node $n$, we use $P^n_{k}=\{p^n_{k1}, p^n_{k2}, \cdots, p^n_{k|P^n_{k}|}\}$ to denote the set of paths of length $k$ adjacent to node $n$, where each path $p$ is a set including the nodes and edges on that path. Now we can define the multidimensional resource connectivity of path $p$ as
\begin{equation}
	 \mathcal{C}_p(t) =[\min_{e\in p}\{r^e_1-u^e_1(t)\}].
\end{equation}

Then we can represent the average connectivity of all $k$-hop paths adjacent to node $n$ as follows:
\begin{equation}
	\mathcal{P}^n_k(t) = \left\{
	\begin{aligned}
		&0, &k=0,\\
		&\frac{\sum_{p\in P^n_{k}}\mathcal{C}_p(t)}{|P^n_{k}|}, &k>0.
	\end{aligned}
	\right.
\end{equation}

For convenience, we subsequently call $	\mathcal{P}^n_{k}$ the $k$-hop path connectivity of node $n$. 

To analyze the resource connectivity of the $k$-hop neighbor nodes of node $n$, we denote the set of neighbor nodes $k$ hops away from node $n$ by $B^n_{k}$. Similar to the path connectivity, we represent the residual multidimensional resource capacity of the neighbor node $b\in B^n_{k}$ in terms of its connectivity by
\begin{equation}
	\mathcal{C}_b(t) = [r^b_1 - u^b_1(t), r^b_2 - u^b_2(t)].
\end{equation}

Then we represent the average connectivity of all $k$-hop neighbor nodes of node $n$ as follows:
\begin{equation}
	\mathcal{B}^n_k(t)  = \left\{
	\begin{aligned}
		&\mathcal{C}_n(t), &k=0,\\
		&\frac{\sum_{b\in B^n_{k}}\mathcal{C}_b(t)}{|B^n_{k}|}, &k>0.
	\end{aligned}
	\right.
\end{equation}

Before proceeding to the next step, an explanation is necessary as to why we calculate connectivity directly using resource values rather than resource utilization rates. This is because there is a hidden precondition for using resource utilization rates to describe the balance degree of the system: the same-dimensional resource capacity of each element in the system is identical. In order to make the proposed fragment model compatible with heterogeneous NFV networks, we compute the connectivity directly using the resource values.

Next, we splice the connectivity of neighboring paths and nodes into a vector and denote it as
\begin{equation}
	\eta^n_{k} = [\mathcal{B}^n_{k}, \mathcal{P}^n_{k}]=[\eta^n_{k 1}, \eta^n_{k 2},\cdots, \eta^n_{k(R_1+R_2)}],
\end{equation}
where $R_1$ denotes the number of node resource dimensions and $R_2$ denotes the number of link resource dimensions. Since each element of this vector represents a certain resource connectivity of node $n$, the $i$-th dimensional resource fragment model of node $n$ can be given by
\begin{equation}
	f^n_{k i} = \frac{1}{\eta^n_{k i} + 1}.
\end{equation}

We can see that the fragment model reaches a maximum value of $1$ when the resource connectivity is $0$.

\section{Problem Formulation}
In this section, we first analyze the optimization variables for the VNF migration problem. Then we define two optimization objectives, which are maximum weighted fragmentation level and migration loss. Finally, we formulate the VNF migration problem.

We first identify the variables of the VNF migration problem with dynamic resources. This problem can be decomposed into three sub-problems: the VNF selection problem, the destination selection problem, and the destination routing problem. The VNF selection problem refers to the choice of which VNFs to migrate at moment $t$; the destination selection problem refers to the choice of which network node to use as the migration destination for a given VNF $v$; the destination routing problem refers to the choice of which paths to use as the new paths for the upstream and downstream VNF links of a given VNF $v$ that has already migrated to node $n$. For the first and second sub-problem, we use a binary variable $x^{n_sn_d}_{v}$ to indicate whether to migrate the VNF $v$ from the source node $n_s$ to the destination node $n_d$; for the third sub-problem, we use a binary variable $y^{p_op_n}_l$ to indicate whether to migrate the VNF link $l$ from the old path $p_o$ to the new path $p_n$. Note that the old path $p_o$ in the superscript of the variable is determined at the current moment. Based on these variables, we can denote the set of VNFs $V^n$ on node $n$ and the set of VNF links $L^e$ on link $e$, respectively, as follows:
\begin{equation}
\begin{aligned}
		V^n(t)=&\{V^n(t-1)\cup \{v|x^{n'n}_v=1, \forall n'\in N\backslash\{n\}\}\}\\&
		\backslash\{v|x^{nn'}_v=1, \forall n'\in N\backslash\{n\}\},
\end{aligned}
\end{equation}
\begin{equation}
	\begin{aligned}
		L^e(t)=&\{L^e(t-1)\cup \{l|y^{p_op_n}_l=1, \forall e\in p_n\}\}\\&
		\backslash\{l|y^{p_op_n}_l=1, \forall e \in p_o \backslash p_n
		\},
	\end{aligned}
\end{equation}
where both $V^n(0)$ and $L^e(0)$ are empty sets when $t=0$.

In order to reduce system overload and migration loss in dynamic networks, the objectives of VNF migration in this paper are to minimize the maximum weighted fragmentation level of all nodes, which will be derived subsequently, as well as to minimize the migration loss. We now introduce these two optimization objectives:
\begin{enumerate}
	\item Maximum Weighted Fragmentation Level
	
	To obtain maximum weighted fragmentation level, we first transform the fragment model based on the total utilization rate of each type of resource in the network. The weight factor $\alpha_i$ for type $i$ resource can be obtained as follows:
	\begin{equation}
		\alpha_i = \left\{
		\begin{aligned}
			\frac{\sum_{n\in N}u^n_i(t)}{\sum_{n\in N}r^n_i}, i \leq R_1,\\
			\frac{\sum_{e\in E}u^e_i(t)}{\sum_{e\in E}r^e_i}, i > R_1.
		\end{aligned}
		\right.
	\end{equation}

	By weighting, we obtain the $k$-th order fragmentation level for each node $n$:
	\begin{equation}
		\mathsf{f}^n_k = \frac{\sum^{R_1+R_2}_{i=1}\alpha_i f^n_{ki}}{\sum^{R_1+R_2}_{i=1} \alpha_i}.
	\end{equation}

	Next, we weight the fragmentation level of node with different receptive fields. Since there exists no criterion for evaluating the optimal value of the receptive field weights, we use a finite geometric series to define the receptive field weights as follows:
	\begin{equation}
		\beta_k =q^k,	0<q\leq1,  k = 0,1,\cdots,K
	\end{equation}
	where $q$ is the weight ratio between the receptive fields of adjacent orders, and $K$ is the upper limit of the receptive field. 
	
	Based on these weights, we define the overall fragmentation level $\mathsf{f}^n$ of node $n$ as follows:
	\begin{equation}
		\mathsf{f}^n = \frac{\sum^{K}_{k=0}\beta_k\mathsf{f}^n_k}{\sum^{K}_{k=0}\beta_k}.
	\end{equation}

	Then we can denote the maximum weighted fragmentation level as
	\begin{equation}
		\mathcal{F}(t)=\max_i{\mathsf{f}^{n_i}}.
	\end{equation}
	\item Migration Cost
	
	To determine the migration loss, we first consider the data transmission time required for the migration. In this paper, we assume that the control plane of the infrastructure network reserves a dedicated bandwidth $BW$ for the migration of each VNF, and take the current memory requirement of the VNF to be migrated as the migration object. Thus the migration transmission time of VNF $v$ can be denoted as follows:
	\begin{equation}
		t^v_{tr} = \frac{r^v_2(t)}{BW}.
	\end{equation}

We then specify that the VNF $v$ migrates to the destination node via the shortest path $p^v_{min}$, and we can calculate the link propagation time for the migration as
\begin{equation}
	t^v_{pr}=\sum_{e\in p_{min}}d_e.
\end{equation}

We can now define the migration loss of a VNF $v$ as the product of the bandwidth requirement of the upstream link of VNF $v$ at moment $t$ and the migration time of VNF $v$:
\begin{equation}\small
	\mathcal{L}^v(t)=\sum_{v'\in V^s\backslash\{v\} }(t^v_{tr}+t^v_{pr})r^{l_{v'v}}_{1}(t)\sum_{n_s\in N}\sum_{n_d\in N}x^{n_sn_d}_v,
\end{equation}
where $V^s$ is the set of VNFs of the SFC $s$, which contains VNF $v$. Next we can define the migration loss for all VNFs as follows:
\begin{equation}\small
	\mathcal{L}(t)=\sum_{s\in S}\sum_{v\in V^s}\mathcal{L}^v(t).
\end{equation}
\end{enumerate}

After obtaining the two optimization objectives, we can give the formulation of the VNF migration problem. We transform the two-objective optimization problem into a single-objective optimization problem by weighting and model it as an integer nonlinear programming problem as follows:
\begin{equation}\small
	\mathcal{P}: \min{\gamma}\sum_{t\in T}\mathcal{F}(t)+(1-\gamma)\sum_{t\in T}\mathcal{L}(t),
\end{equation}
\begin{subequations}
\begin{flalign}\small
	\textit{s.t.}\quad&u^n_i(t)\leq\rho r^n_i, \forall n\in N, t\in T, i\in\mathcal{R}_1,\\
&u^e_i(t)\leq\rho r^e_i, \forall e\in E, t\in T, i\in\mathcal{R}_2,\\
&d^p_{v_2}+\!\!\!\sum_{p_n\in P_{v_1v_2}}\sum_{e\in p_n} d^e \gamma^{p_op_n}_{l_{v_1v_2}}\leq D_{l_{v_1v_2}}, \\
&\forall v_1,v_2\in V^s, s\in S, \nonumber\\
&\sum_{n_s\in N}\sum_{n_d\in N}x^{n_sn_d}_v\leq 1,\forall v\in V^s, \\
&\sum_{p_n\in P_{v_1v_2}}y^{p_op_n}_{l_{v_1v_2}}\leq 1,\forall v_1,v_2\in V^s, s\in S,\\
&x^{n_sn_d}_{v_1}, y^{p_op_n}_{l_{v_1v_2}}\in \{0,1\}, \forall v_1,v_2\in V^s, s\in S, \\
 & n_s, n_d\in N,\nonumber\\
& r^n_i, r^e_i, r^v_i(t), r^l_j(t)\geq 0,\forall n\in N,e\in E,\\ 
& v\in V^s, l\in L^s, s\in S, \in\mathcal{R}_1, j\in\mathcal{R}_2,\nonumber
\end{flalign}
\end{subequations}
where $\rho$ denotes the overload threshold, $\mathcal{R}_1$ denotes the set of node resource type indexes, $\mathcal{R}_2$ denotes the set of link resource type indexes, and $P_{v_1v_2}$ denotes the set of paths between VNFs $v_1$ and $v_2$. Constraints (21a) and (21b) ensure that the infrastructure resource utilization does not exceed the resource limit. Constraint (21c) guarantees that the latency of the VNF does not exceed the latency limit. Constraints (21d) and (21e) guarantee that a VNF can only be migrated to a single node and that a VNF link can only be migrated to a single path. Constraints (21f) and (21g) limit the range of variables and constants.

This problem can be regarded as a variant of the bin-packing problem, which is NP-hard and cannot be solved in polynomial time. Since VNF migration requires rapid adjustment of resource allocation based on traffic changes, the large amount of time required to find an exact solution is usually unacceptable. In this paper, we propose a deep learning approach to obtain sub-optimal solutions quickly in polynomial time. With a unique neural network design and pre-training, the proposed method is able to effectively capture the fragmentation level of the network and reduce network overload and migration loss.
\vspace{-3mm}
\section{Methodology}
In this section, we first present the overall framework of the proposed migration strategy in Sec. \ref{framework}. In Sec. \ref{model_architecture}, we give the details of the proposed neural network model. Finally, we present the dataset generation and neural network training in Sec. \ref{dataset} and Sec. \ref{train}, respectively.
\vspace{-3mm}
\subsection{Framework}\label{framework}
\begin{figure*}[!t]
	\centering
	\includegraphics[width=6.0in]{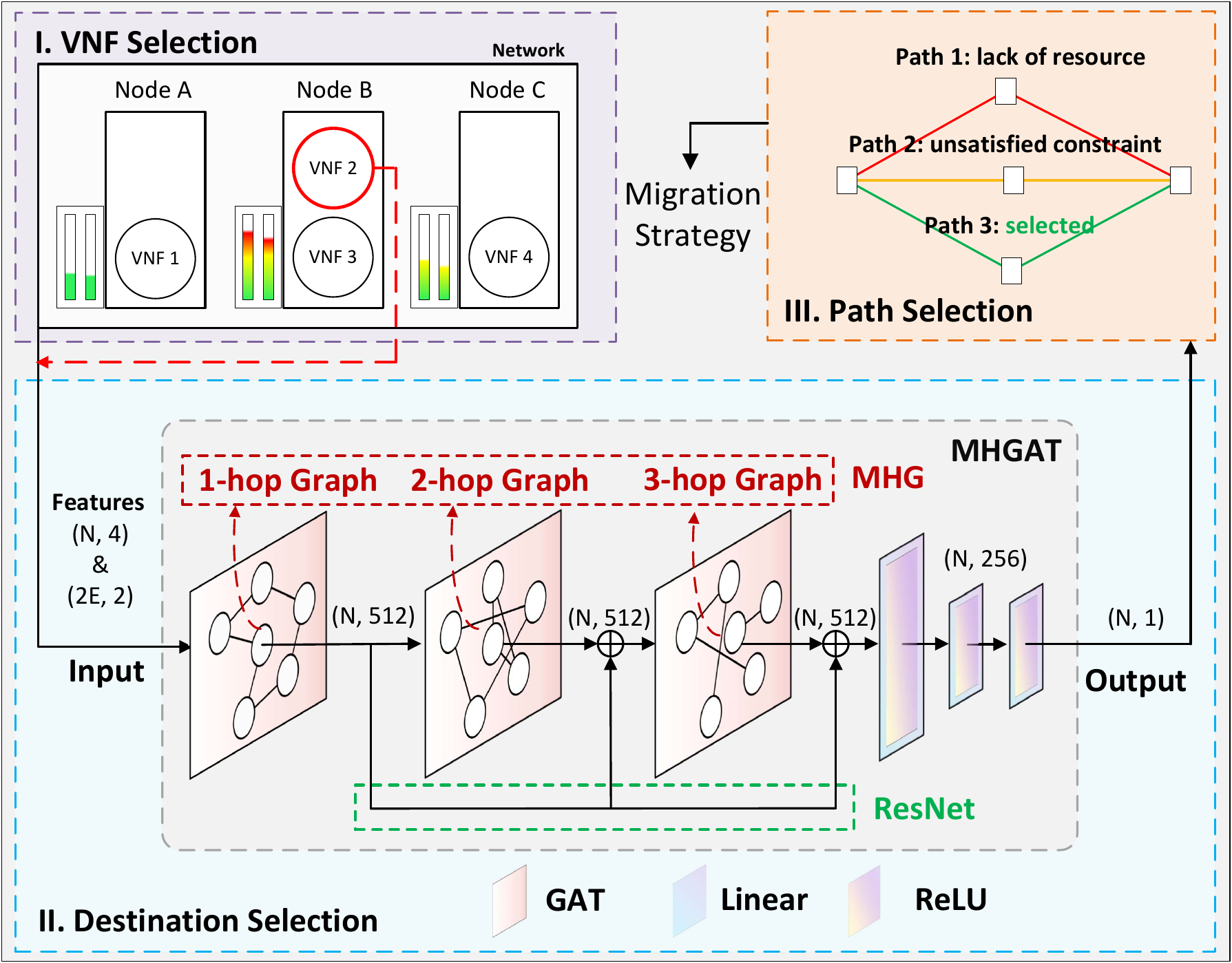}
	\caption{Overall framework of the proposed MHGAT method.}
	\label{fig_framework}
\end{figure*}
As shown in Fig. \ref{fig_framework}, the framework of the proposed MHGAT method consists of three main components which are VNF selection, destination selection and path selection. We now specify these three components as follows.
\begin{enumerate}
	\item VNF Selection
	
	In the event that a node overload failure occurs and triggers the migration method to run, the migration time $(t_{tr}+t_{pr})$ needs to be minimized in order to minimize the loss due to VNF migration. We first consider the transmission time $t_{tr}$. Although it is determined by the memory requirement of the VNF, selecting the VNF with the least memory requirement has no effect on intervening in an overload where the overloaded resource is not memory and can lead to persistent migration. Thus, to handle an overload, we consider the overloaded resource type. We denote the overloaded resource type index by $j$ and select the VNF to be migrated in the overloaded node $n$ by the following equation:
\begin{equation}\small
	\label{eq_vnf_select}
	v \!\!=\!\! \left\{
		\begin{aligned}
			&\sideset{}{r^v_j(t)}{\arg\max}_{v\in V^n},\quad u^n_j(t)\!\!-\!\!\max_{v\in V^n}{\!r^v_j(t)}\!>\!\rho r^n_j,\\
			&\sideset{}{\{r^v_j(t)|u^n_j(t)-r^v_j(t)\leq \rho r^n_j\}}{\arg\min}_{v\in V^n},\quad
			else.
		\end{aligned}
		\right.
\end{equation}

When $u^n_j(t)\!\!-\!\!\max_{v\in V^n}{\!r^v_j(t)}\!>\!\rho r^n_j$, there is no VNF that can eliminate the overload by one-time migration, and then the VNF that requires the most overloaded resource is selected as the VNF to be migrated. Otherwise, there exists VNF that can eliminate the overload after being migrated. In this case, we select the VNF that can eliminate the overload after migration and requires the least overloaded resource as the VNF to be migrated. In the latter case, the action logic is to minimize the resource pressure delivered to other nodes on the premise of eliminating overload.

	\item Destination Selection
	
	After selecting the VNFs to be migrated, we perform destination selection based on a deep learning method in order to reduce the maximum weighted fragmentation level in the network after migration and to quickly generate a solution. We first introduce the principle of the deep learning method. Then we give the inputs and outputs of the neural network, whose specific design is presented in Sec. \ref{model_architecture}.

 Conventional heuristic algorithms, such as the greedy algorithm, usually only consider migrating the VNF to the node with the largest remaining capacity of the overloaded resources, which may not be able to alleviate the fragmentation phenomenon; whereas the exhaustive search method needs to simulate the migration and obtain the optimal solution based on the optimization objective value after the simulated migration, which consumes a large amount of computational time. In this paper, we design a deep learning method with the goal of minimizing the maximum weighted fragmentation level after migration, which skips the simulated migration computation and achieves fast destination selection by capturing the multidimensional resource characteristics of the network through the improved GAT and directly outputting the approximate maximum weighted fragmentation level information after migration.
	
	In order for the neural network to output information about the network state after migration based on the existing information, the input data needs to contain node resource information, link resource information and information about the VNF $v$ to be migrated. We splice the network node information and VNF node resource requirement information and denote them by tensor $X_v\in \mathbb{R}^{|N|\times2R_1 }$:
	\begin{equation}
		\mathcal{X}_v=\left[
		\begin{matrix}
			\frac{r^v_1(t)}{r^{n_1}_1}&\frac{r^v_2(t)}{r^{n_1}_2}&\frac{u^{n_1}_1(t)}{r^{n_1}_1}&\frac{u^{n_1}_2(t)}{r^{n_1}_2}\\
			\frac{r^v_1(t)}{r^{n_2}_1}&\frac{r^v_2(t)}{r^{n_2}_2}&\frac{u^{n_2}_1(t)}{r^{n_2}_1}&\frac{u^{n_2}_2(t)}{r^{n_2}_2}
			\\
			&\vdots&&
			\\
			\frac{r^v_1(t)}{r^{n_{|N|}}_1}&\frac{r^v_2(t)}{r^{n_{|N|}}_2}&\frac{u^{n_{|N|}}_1(t)}{r^{n_{|N|}}_1}&\frac{u^{n_{|N|}}_2(t)}{r^{n_{|N|}}_2}
		\end{matrix}\right],
	\end{equation}
where row $i$ of the tensor represents the multidimensional resource requirement information of the VNF
 to be migrated and the multidimensional resource information of the node $n_i$, the first $R_1$ columns are used to indicate the requirement resource occupancy ratio of the VNF, and the last $R_1$ columns are used to indicate the current resource utilization of the node. 
 
 Next, we need to represent the resource requirement information of the VNF links connected to the VNF  to be migrated and hardware link resource information. We use the average of the bandwidth requirements of all the VNF links connected to the VNF to be migrated as the link resource requirement information of the VNF and denote it as $\ell^v_i$ as follows:
 \begin{equation}
 	\ell^v_i = \sum_{v'\in V^s\backslash\{v\}}r^{l_{v'v}}_i(t) +\sum_{v'\in V^s\backslash\{v\}}r^{l_{vv'}}_i(t),
 \end{equation}
 where the subscript $i$ denotes the index of the link resource type. Then we can denote by tensor $\mathcal{E}_v \in R^{2|E|×2R_2 }$ the network link information and the link resource requirement information of the VNF to be migrated:
 \begin{equation}\small
 	\label{net_edge}
 	\mathcal{E}_v \!\!=\!\!\left[
 	\begin{matrix}
 		\!\frac{\ell^v_1}{r^{e_1}_1} 
 		\!&\!\!\!\!\frac{\ell^v_1}{r^{e_1}_1}
 		\!&\!\!\!\!\frac{\ell^v_1}{r^{e_2}_1}
 		\!&\!\!\!\!\frac{\ell^v_1}{r^{e_2}_1}
 		\!&\!\!\!\!\cdots
 		\!&\!\!\!\!\frac{\ell^v_1}{r^{e_{|E|}}_1}
 		\!&\!\!\!\!\frac{\ell^v_1}{r^{e_{|E|}}_1}
 		\\
 		\frac{u^{e_1}_1(t)}{r^{e_1}_1}
 		\!&\!\!\!\!\frac{u^{e_1}_1(t)}{r^{e_1}_1}
 		\!&\!\!\!\!\frac{u^{e_2}_1(t)}{r^{e_2}_1}
 		\!&\!\!\!\!\frac{u^{e_2}_1(t)}{r^{e_2}_1}
 		
 		\!&\!\!\!\!\cdots
 		\!&\!\!\!\!\frac{u^{e_{|E|}}_1\!(t)}{r^{e_{|E|}}_1}
 		\!&\!\!\!\!\frac{u^{e_{|E|}}_1\!(t)}{r^{e_{|E|}}_1}
 	\end{matrix}
 	\right]^T,
 \end{equation}
where the first $R_2$ rows of each column indicate the link resource requirement information of the VNF, and the last $R_2$ rows indicate the current resource utilization of the network links. Since the neural network needs to process the edges of the graph using directed edges as a criterion, the information for each edge needs to be repeated twice. In addition, for ease of representation, we use matrix transposition in Eq. (\ref{net_edge}).

Now we can summarize that the inputs of the neural network include node information $\mathcal{X}_v$ and link information $\mathcal{E}_v$. Since our goal is to obtain the maximum weighted fragmentation level after migration through the neural network, the output information should include the weighted fragmentation level of each node after migration, i.e., the output information is a tensor of size $|N|\times 1$, denoted as $\mathcal{O}_v$.

	\item Path Selection
	
	Path selection is triggered by VNF migration or link overload. Since the network graph is invariant, we reduce the time complexity of the VNF migration method by searching the paths between each node in advance through depth-first search. Therefore, when triggering path selection, the shortest path in the finite set of paths that satisfies the delay and bandwidth constraints can be directly selected, while the time complexity depends on the number of paths searched in advance.
\end{enumerate}
\vspace{-3mm}
\subsection{Model Architecture}\label{model_architecture}
The design of our neural network model architecture is based on GAT. GAT focuses on a single node in the network and extracts node features by calculating its similarity coefficients with its neighbors, which is similar to our idea of calculating fragmentation level. In GAT, the similarity coefficient $e_ij$ of nodes $n_i$ and $n_j$ is calculated as follows:
\begin{equation}
	e_{ij}=a(\mathbf{W}\mathcal{X}_v[i]||\mathbf{W}\mathcal{X}_v[j]),
\end{equation}
where $\mathbf{W}$ is a weight matrix shared by the nodes, $a(\cdot)$ is a parameter matrix used to map feature dimension to output dimension, and the symbol $||$ denotes the splicing of two tensors into a single tensor. In the case of considering edge features, the coefficient can be rewritten as
\begin{equation}
	e_{ij} = a(\mathbf{W}\mathcal{X}_v[i]||\mathbf{W}\mathcal{X}_v[j])+a_2(\mathcal{E}_{v,ij}),
\end{equation}
where $a_2(\cdot)$ is the parameter matrix used to map the dimension of the edge features. In order to make the similarity coefficients easily comparable across nodes, GAT normalizes them using the $softmax$ function and activates the similarity coefficients non-linearly using the $LeakyReLU$ function:
\begin{equation}
	\alpha_{ij} = \frac{exp(LeakyReLU(e_{ij}))}{\sum_{k=1}^{|N|}exp(LeakyReLU(e_{ik}))},
\end{equation}
where the $softmax$ and $LeakyReLU$ functions are commonly used activation functions in the field of deep learning to activate neurons in neural network. Finally the output feature of node $n_i$ is obtained by weighting and activating the similarity coefficients:
\begin{equation}\small
	\label{net_output_single}
	h_i = \sigma(\sum_{j=1}^{|N|}\alpha_{ij}\mathbf{W}\mathcal{X}_v[j]),
\end{equation}
where $\sigma(\cdot)$ denotes the activation function. In order to stabilize the learning process, GAT introduces the multi-head attention mechanism, where each head feature independently performs the operation of Eq. (\ref{net_output_single}) and finally concatenates the multi-head features as follows:
\begin{equation}\small
	h_i=concat^M_{m=1}(\sigma(\sum_{j=1}^{|N|}\alpha^m_{ij}\mathbf{W}^m\mathcal{X}_v[j])),
\end{equation}
where $concat^M_{m=1}(\cdot)$ denotes the splicing tensor function, $m$ denotes the index of heads and $M$ denotes the total number of head.  Now we have obtained a complete GAT layer considering edge features. The neural network structure design is presented next.

As shown in Fig. \ref{fig_framework}, our proposed multi-hop GAT (MHGAT) consists of three special GAT layers and three linear layers. Note that the number of GAT layers is related to the maximum receptive field $K$, which is $K + 1$. In this section and subsequent sections, the maximum receptive field is empirically set to $2$. The first GAT layer of MHGAT accepts the input tensor and the original graph topology $G_1$ to get the output $\mathcal{O}_1\in \mathbb{R}^{|N|\times\theta_1}$:
\begin{equation}\small
	\mathcal{O}_1=GATlayer(\mathcal{X}_v,\mathcal{E}_v, G_1).
\end{equation}
Starting from the second GAT layer, we input a different graph topology, i.e., the $2$-hop graph topology. In a $n$-hop graph, edges between each pair of nodes can only exist if there is a $n$-hop path between that pair of nodes. With the $n$-hop graph topology, the GAT layer can capture the similarity coefficients between pairs of nodes with a distance of $n$ more directly, thus obtaining a wider range of node-related information. This is why the number of GAT layers is $K+1$, since the receptive field features we need to capture are counted from $0$ to $K$. The output of the second GAT layer can be represented as follows:
\begin{equation}
	\mathcal{O}_2=GATlayer(\mathcal{O}_1, G_2),
\end{equation}
where $G_2$ is the $2$-hop graph topology. In order not to lose the output information of the first GAT layer, we utilize the residual network (ResNet), a neural network structure used to solve the deep network degradation problem, to connect the output information of the first GAT layer to the second GAT layer:
\begin{equation}\small
	\mathcal{O}'_2 = \mathcal{O}_1+\mathcal{O}_2.
\end{equation}
With the ResNet, the output information of the first GAT layer can be retained in the subsequent layers. Then $\mathcal{O}_2$ is input into the third GAT layer to obtain the 3-hop graph features:
\begin{equation}\small
	\mathcal{O}_3=GATlayer(\mathcal{O}'_2, G_3).
\end{equation}
Similarly, we utilize the ResNet to retain the original graph information as follows:
\begin{equation}\small
	\mathcal{O}'_3 = \mathcal{O}_1+\mathcal{O}_3.
\end{equation}
Finally, we integrate the feature information using three linear layers connected to the activation function and obtain the output: 
\begin{equation}\small
	\mathcal{O}_v=ReLU(L_3(ReLU(L_2(ReLU(L_1(\mathcal{O}'_3)))))).
\end{equation}

\subsection{Dataset Generation}\label{dataset}
This subsection introduces how to generate the deep learning training dataset we need. Unlike reinforcement learning, which generates data by interacting with the environment, training a neural network based on a pre-generated dataset allows the network to converge quickly and achieve accurate outputs.

The goal of the neural network is to obtain the weighted fragmentation level after migration, and the dataset can achieve this goal in advance via the exhaustive method. We do not run any VNF migration algorithm and deploy VNFs into the network by the generic VNF deployment algorithm until there is an overload. At this point we run the VNF selection algorithm to determine the VNF $v$ to be migrated, followed by migrating it to $|N|$ nodes in $|N|$ simulated networks respectively and obtaining the corresponding maximum weighted fragmentation levels. Thus the training data label can be represented as follows:
\begin{equation}\small
	\hat{\mathcal{O}}_v=\left[
	\begin{matrix}
		\max_i{\{\mathsf{f}^{n_i}|x^{n_sn_1}_v=1\}}
		\\
		\max_i{\{\mathsf{f}^{n_i}|x^{n_sn_2}_v=1\}}
		\\
		\vdots
		\\
		\max_i{\{\mathsf{f}^{n_i}|x^{n_sn_{|\!N\!|}}_v=1\}}
	\end{matrix}
	\right],
\end{equation}
where $n_s$ denotes the node where VNF $v$ was located before migration. We repeat the above process a large number of times to form the training dataset.
\subsection{Neural Network Training}\label{train}
Based on the training dataset generated in Sec. \ref{dataset}, we use the mean square error (MSE) function and the Adam optimizer, which are commonly used in the field of deep learning to train neural network models. We feed the input data from the training set into the neural network model to get the raw output and calculate the MSE between it and the training set labels, as follows:
\begin{equation}\small
	\mathbb{L} = MSE(\mathcal{O}_v - \hat{\mathcal{O}}_v),
\end{equation}
where $\mathbb{L}$ represents the loss function of the deep learning method. After the training in the dataset is completed, MHGAT can obtain the approximate maximum weighted fragmentation level after migrating VNFs to different nodes online based on the current network state and the information of the VNF to be migrated.
\begin{algorithm}[!t]
	\caption{MHGAT Migration Method}
	\label{alg_MHGAT}
	\begin{algorithmic}[1]
		\STATE \textbf{Input:} $G$, $S$, trained MHGAT model.
		\IF{$\exists n\in N, i=1,\cdots, R_1, u^n_i(t)>\rho r^n_i$}
		\FOR{$n\in N_{over}$}
		\STATE $loop,blacklist\leftarrow 0, \emptyset$.
		\WHILE{$\exists i, u^n_i(t)>\rho r^n_i $\AND $loop<\zeta_l$}
		\STATE $loop\leftarrow loop +1$.
		\STATE Find resource type $j=\max_j{\frac{u^n_j(t)}{r^n_j}}$.
		\FOR{$v\in V^n$}
		\IF{$v\in blacklist$}
		\STATE \textbf{Continue}.
		\ENDIF
		\STATE Select VNF $\hat{v}$ according to Eq. (\ref{eq_vnf_select}).
		\ENDFOR
		\STATE  $\mathcal{O}_{\hat{v}}\leftarrow MHGAT(\mathcal{X}_{\hat{v}}, \mathcal{E}_{\hat{v}})$.
		\STATE Select the node with the maximum value in $\mathcal{O}_{\hat{v}}$ that satisfies the constraints as destination.
		\IF{Migration failes}
		\STATE $blacklist\leftarrow blacklist\cup\{\hat{v}\}$
		\ENDIF
		\ENDWHILE
		\ENDFOR
		\ENDIF
		\IF {$\exists e\in E, i=1,\cdots, R_2, u^e_i(t)>\rho r^e_i$}
		\FOR{$e\in E_{over}$}
		\STATE Find resource type $j=\max_j{\frac{u^e_j(t)}{r^e_j}}$.
		\STATE $\Delta\leftarrow u^e_j(t) - \rho r^e_j$.
		\FOR{$l\in L^e$}
		\IF{$\Delta\leq 0$}
		\STATE \textbf{Break}.
		\ENDIF
		\STATE Migrate link $l$ to the shortest path that does not include $e$.
		\STATE $\Delta\leftarrow \Delta - r^l_j(t)$.
		\ENDFOR
		\ENDFOR
		\ENDIF
	\end{algorithmic}
\end{algorithm}
\subsection{Algorithm Description}
In this subsection, we present the algorithmic details of the MHGAT method framework in Fig. \ref{fig_framework}. The main process of our proposed MHGAT migration method is shown in Algorithm \ref{alg_MHGAT}. Lines 2-21 of the algorithm are used to eliminate node overload while lines 22-34 are used to eliminate link overload. We iterate through each overloaded node in lines 3-20. In line 4 we initialize the single node migration counter and VNF blacklist. VNFs are migrated out of the current node recurrently when the current node is overloaded and the counter does not reach the specified upper limit $\zeta_l$. If the migration fails, the VNF is added to the blacklist. In lines 23-33, we traverse each overloaded link and recurrently migrate VNF links from the set of VNF links of the network link to the shortest path that satisfies the constraints and does not pass through the network link.
\vspace{-3mm}
\subsection{Complexity Analysis}
In this subsection we calculate the time complexity of the MHGAT method. We first calculate the complexity of the first 21 lines of Algorithm \ref{alg_MHGAT}, which can be expressed as $O(|N|(|V|+|N|+|E|))$, where the loop in line 3 contributes $O(|N|)$ and the loop in line 8 contributes $O(|V|)$. Note that the $V$ denotes the set of VNFs for all SFCs. In line 14, obtaining the neural network inputs $\mathcal{X}_{\hat{v}}$ and $\mathcal{E}_{\hat{v}}$ contributes complexity $O(|N|+|E|)$, while outputting a result contributes complexity $O(\theta_1|N|+\theta_2|E|)$, where $\theta_1$ and $\theta_2$ are related to the number of neural network parameters. Thus the total complexity of the first 21 lines can be calculated as $O(|N|(|V|+|N|+|E|))$. We then compute the complexity of lines 22-34. Since the paths between nodes are searched in advance in this paper, the complexity calculation does not consider the path search. Thus we can easily get the complexity of lines 22-34 as $O(|E||L|)$, where $L$ denotes the set of VNF links of all SFCs. Finally, we can get the complexity of MHGAT method as $O(|N||V|+|N|^2+|N||E|+|E||L|)$.
\section{Performance Evaluation}
In this section, we verify the effectiveness of our work. We first present the simulation setup and the benchmark migration algorithm used to compare the performance. Then, we verify the correlation between the proposed maximum weighted fragmentation level and the ratio of overloads in the time period. The effectiveness of each module of the designed MHGAT neural network is also validated through ablation experiments. Finally we analyze the performance of the MHGAT method. Our simulation code can be accessed online \cite{zhang2024frag}.
\vspace{-3mm}
\subsection{Simulation Setting}\label{sim_set}
Our simulations are mainly performed on the National Science Foundation Network (NSFNET) topology, which has 14 network nodes and 22 network links. Among them, each node has a CPU frequency capacity of 32 GHz and a memory capacity of 64 GB, while each link has a bandwidth capacity of 5 MBps, a propagation delay of 1 to 5 ms, and a migration bandwidth of 1 MBps allocated for each VNF. As for the SFC, each SFC is limited to a latency between 20 ms and 50 ms, with a lifetime of 1 to 100 time slots. In addition, the VNF has a processing delay of 1 to 5 ms. Without additional statements, we set the SFC arrival rate to 10 per time slot, the overload threshold $\rho$ to 0.5, and the optimization objective weighting factor $\gamma$ to 0.9. Finally, we extract the time-varying resource requirements of VNFs and VNF links from the public Bitbrains dataset \cite{bit2015dataset}. All simulations were run on a computer configured with an Intel Core i7-9700 CPU, NVIDIA GeForce RTX 3080 graphics card and 24G RAM. 
\vspace{-3mm}
\subsection{Benchmarks}\label{sim_compare}
We implemented two benchmark algorithms for performance comparison, which are described in detail as follows.
\begin{enumerate}
	\item LBVMC \cite{yao2023an}: LBVMC is a heuristic load balancing algorithm proposed in the latest research on multidimensional resource fragmentation. It is innovative in reducing unnecessary migrations caused by occasional load fluctuations by predicting future loads. The disadvantages of LBVMC compared to our work are the lack of quantitative analysis of the multidimensional resource fragmentation level and the additional computational overhead associated with prediction.
	\item Greedy \cite{xia2016optimized}: The Greedy algorithm is a classical heuristic algorithm. It proceeds by i) selecting the VNF that occupies the most overloaded resource; ii) migrating it to the node with the maximum remaining overloaded resource; and iii) selecting the shortest path for VNF links. The Greedy algorithm is difficult to obtain the global optimal solution, but has strong time efficiency and can produce a migration strategy with acceptable effects in a very short time.
\end{enumerate}
\vspace{-3mm}
\subsection{Validation of Fragmentation Level}
\label{sim_frag}
We verify the correlation between the maximum weighted fragmentation level and the ratio of overloaded time slots to total cycles during the operating cycle in this subsection. We recurrently increase the SFC arrival rate from 1 to 10 and continuously collect data for 1000 running cycles to obtain in-cycle averages of multiple load metrics in the system for different overload rates. 

\begin{figure}[!t]
	\centering
	\vspace{-3mm}
	\subfloat[\scriptsize{Metrics values with different overload ratios.}]
	{
		\includegraphics[width=1.5in]{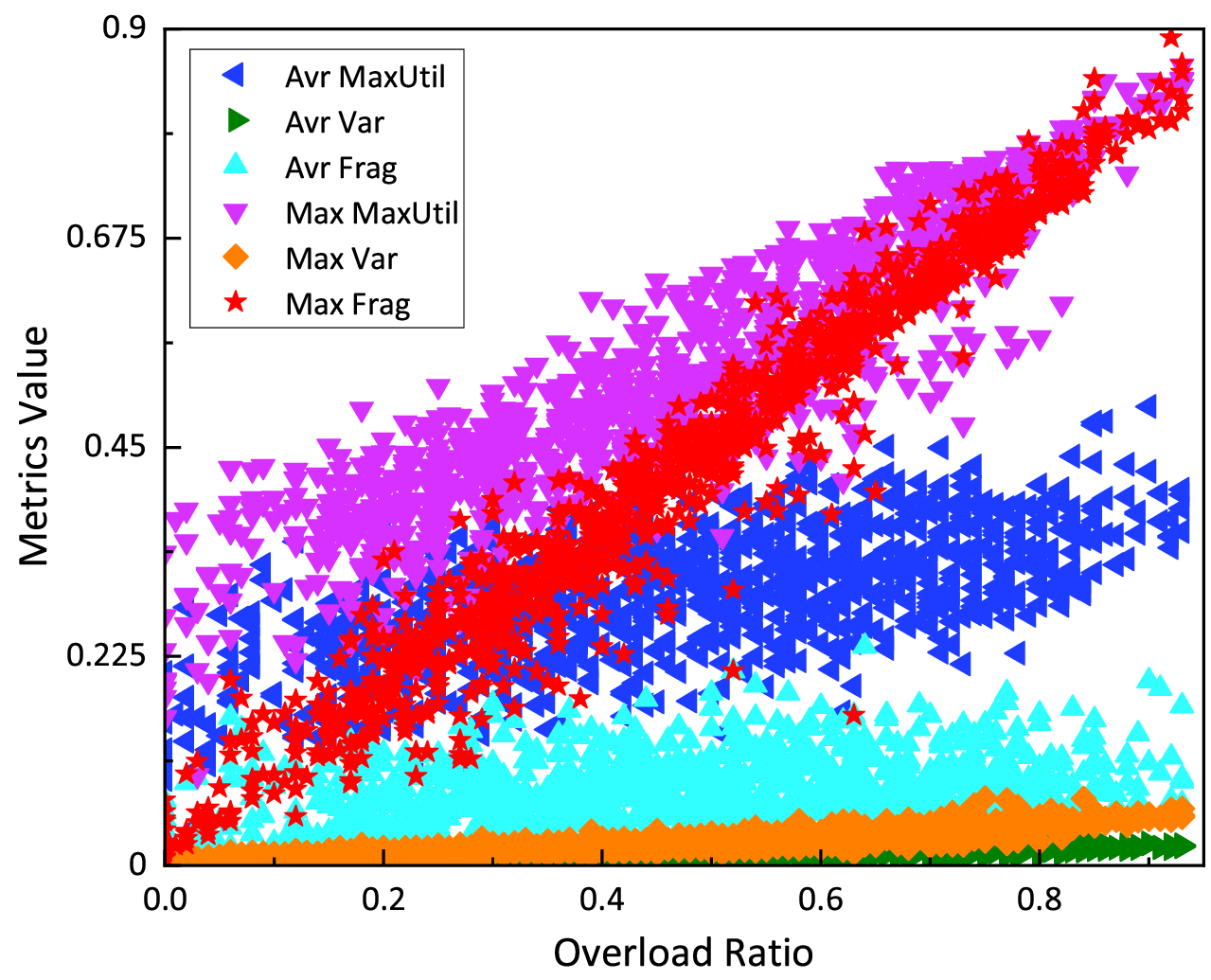}
		\label{fig_frag_val}}
	\hfil
	\subfloat[\scriptsize{Values of different types of correlation coefficients.}]
	{\hspace{2mm}\includegraphics[width=1.4in]{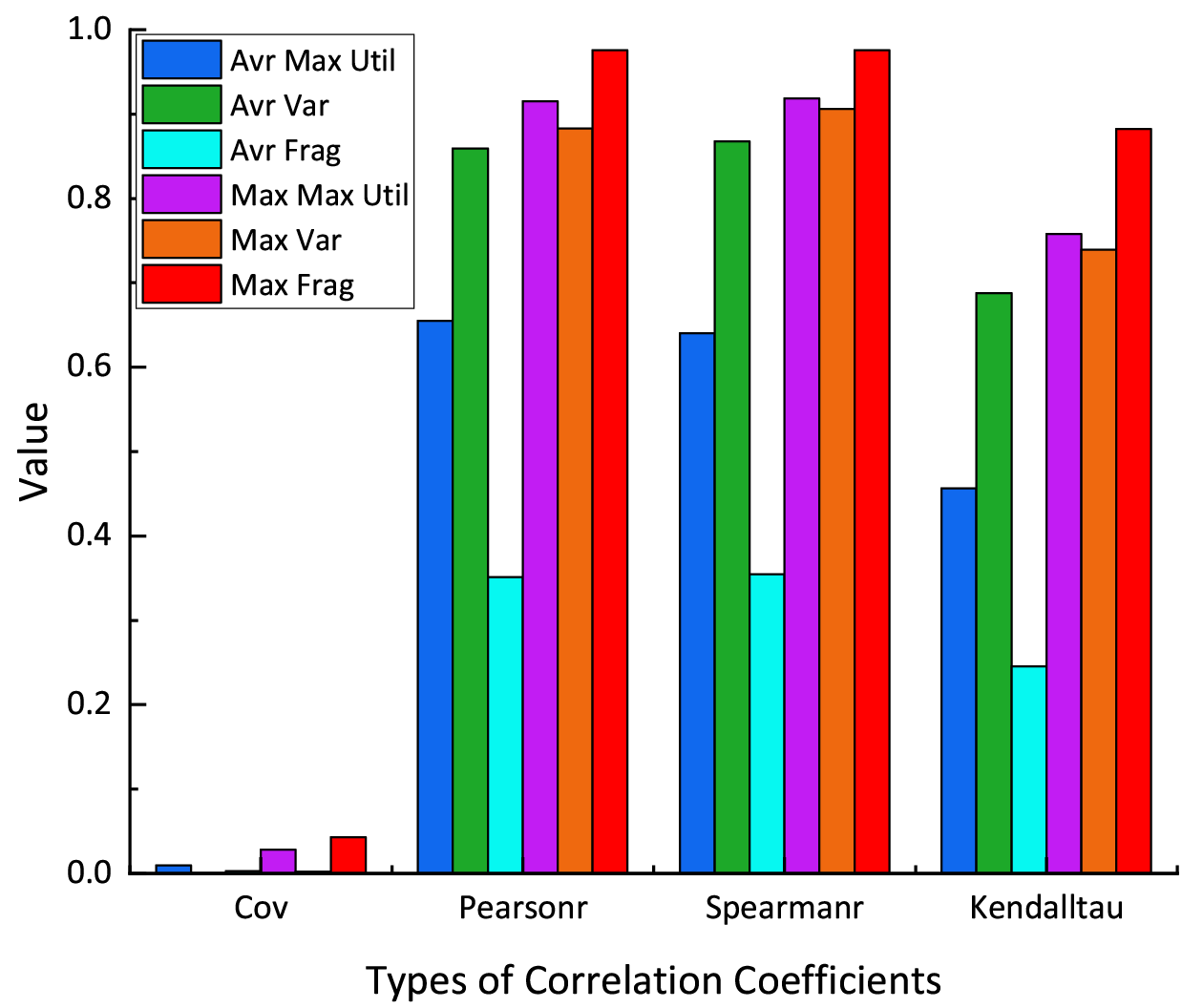}
		\label{fig_corr}}
	\vspace{-2mm}
	\caption{Correlation comparison of different metrics with overload ratio.}
	\label{fig_frag}
\end{figure}

Fig. \ref{fig_frag} shows the correlation comparison of different metrics with overload ratio. In Fig. \ref{fig_frag}(a), we visualize the values of the load metrics with different overload ratios in the form of scatter plots, where the metrics are the maximum uni-dimensional resource utilization averaged across multidimensional resource types (Avr MaxUtil), the average uni-dimensional resource variance across multidimensional resources (Avr Var), the average weighted fragmentation level across multidimensional resources (Avr Frag), the maximum uni-dimensional resource maximum utilization among multidimensional resources (Max MaxUtil), the maximum uni-dimensional resource variance among multidimensional resources (Max Var), and the maximum weighted fragmentation level across multidimensional resources (Max Frag). Fig. \ref{fig_frag}(b) gives the specific values of the four commonly used correlation coefficients. From Fig. \ref{fig_frag} we can see that the maximum value of the proposed weighted fragmentation level has the strongest correlation with the overload ratio, while the maximum value of the maximum utilization ranks second. This is because in each time slot, Max MaxUtil characterizes the system conditions that are prone to overloading, while Max Frag further considers the overload due to the fragmentation phenomena. Specifically, the higher the Max MaxUtil, the scarcer the current uni-dimensional resource, while the higher the Max Frag, the scarcer the current uni-dimensional or multidimensional resource, making it more difficult to improve the system condition after migration. Therefore, Max Frag has a stronger correlation with the overload ratio. In addition, this simulation also verifies the reasonableness of using Max Frag as one of our optimization objectives.
\vspace{-3mm}
\subsection{Validation of MHGAT Network}
\label{sim_net}
In this subsection, we validate the effectiveness of the proposed neural network model by ablation experiments. We compare the complete MHGAT model with the MHGAT model without GAT module, the MHGAT model without ResNet module and the MHGAT model without multi-hop graph (MHG) module, respectively. The neural network convergence performance and MSE of the output results with labels are the evaluation metrics.

\begin{figure}[!t]
	\centering
	\vspace{-3mm}
	\includegraphics[width=3in]{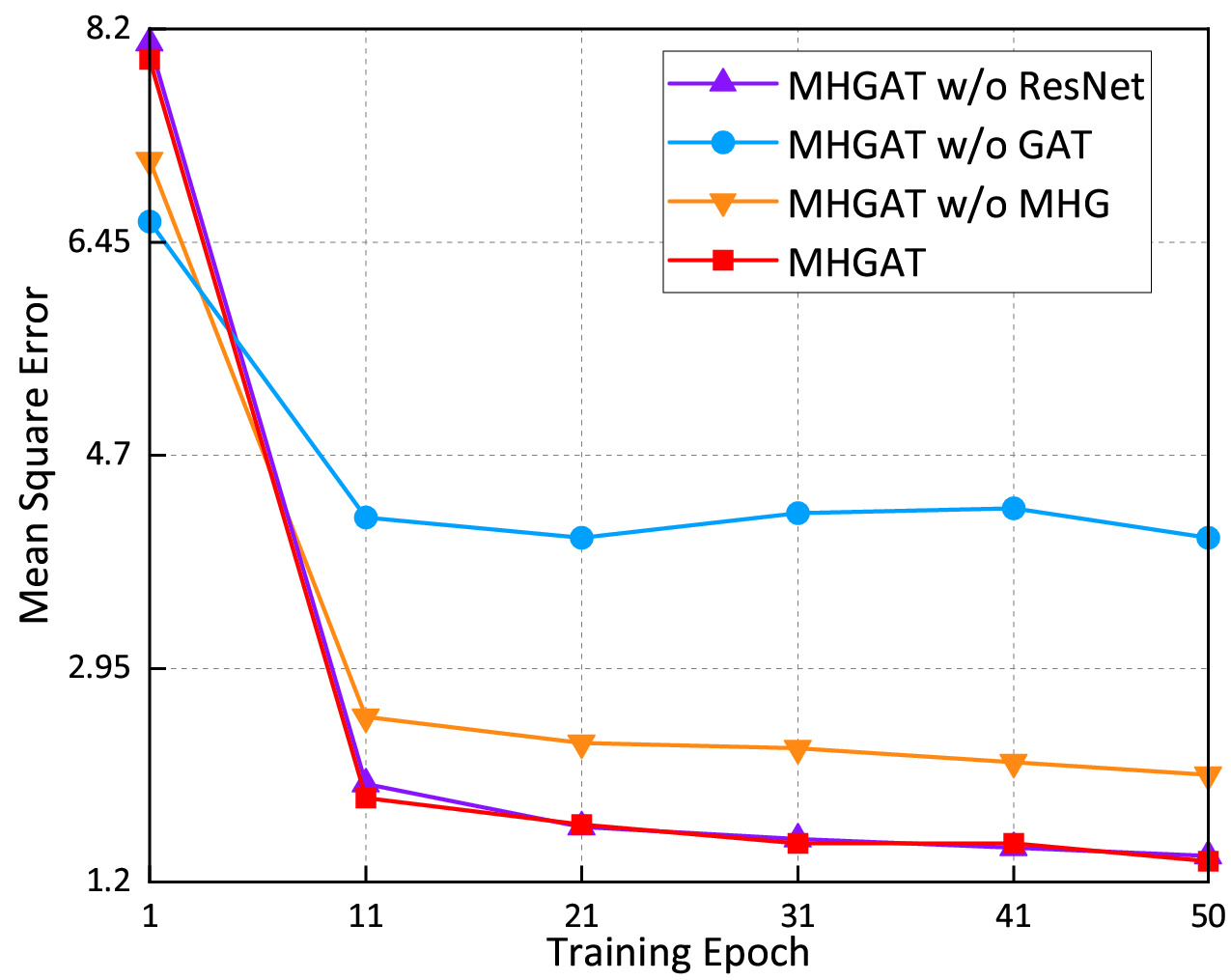}
	\caption{Convergence performance of network models with different training epochs.}
	\label{fig_net}
	\vspace{0mm}
\end{figure}
Fig. \ref{fig_net} illustrates that GAT has the most significant impact on the accuracy of the converged network model, followed by the proposed MHG, while ResNet has the least impact. This is due to the fact that 1) GAT is able to effectively capture the node and edge features of the graph; 2) the proposed MHG effectively captures the node and edge features that are multiple hops apart; and 3) the main role of ResNet is to alleviate the overfitting of the network model by avoiding the loss of information from the lower layers of the network. In addition, in terms of convergence time, the models all converged at similar training epochs (11-21 epochs), which is different from the high convergence episodes in the field of reinforcement learning \cite{liu2023deep, wang2022multi}.

Table \ref{tab_net} gives a comparison of the MSE means and variances for different models with the two real network topologies, where the labeled datasets are randomly generated. The added USbackbone topology has 24 network nodes and 43 network links. From Table \ref{tab_net} we can see that the GAT module has the most significant impact on model performance in both topologies, while the second ranked modules are ResNet and MHG, respectively. The impact of the ResNet module on the network performance is relatively stable across the two topologies, while the MHG module has a greater impact in the USbackbone topology due to the fact that the number of the nodes and edges of the USbackbone topology is higher, which necessitates the capture of multi-hop node features by the MHG module. Thus, the effectiveness of each module of the proposed MHGAT model is validated.
\begin{table}[]
	\caption{Results of ablation experiments on MHGAT model using randomly generated dataset with different topologies.}
	\centering
	\scriptsize
			\label{tab_net}
	\begin{tabular}{@{}|l|ll|ll|@{}}
		\toprule
		Topologies       & \multicolumn{2}{c|}{NSFNET}                              & \multicolumn{2}{c|}{USbackbone}                          \\ \midrule
		Metrics          & \multicolumn{1}{l|}{MSE Mean}         & MSE Var          & \multicolumn{1}{l|}{MSE Mean}         & MSE Var          \\ \midrule
		MHGAT w/o GAT    & \multicolumn{1}{l|}{0.02897}          & 0.01989          & \multicolumn{1}{l|}{0.06127}          & 0.03863          \\
		MHGAT w/o ResNet & \multicolumn{1}{l|}{0.01427}          & 0.00635          & \multicolumn{1}{l|}{0.02424}          & 0.00815          \\
		MHGAT w/o MHG    & \multicolumn{1}{l|}{0.01172}          & 0.00351          & \multicolumn{1}{l|}{0.04722}          & 0.02552          \\
		MHGAT            & \multicolumn{1}{l|}{\textbf{0.01112}} & \textbf{0.00334} & \multicolumn{1}{l|}{\textbf{0.02272}} & \textbf{0.00577} \\ \bottomrule
	\end{tabular}
\end{table}
\begin{figure*}[!t]
	\vspace{-3mm}
	\centering
	\subfloat[\scriptsize{Acceptance rate with different SFC arrival rates.}]
	{\includegraphics[width=1.5in]{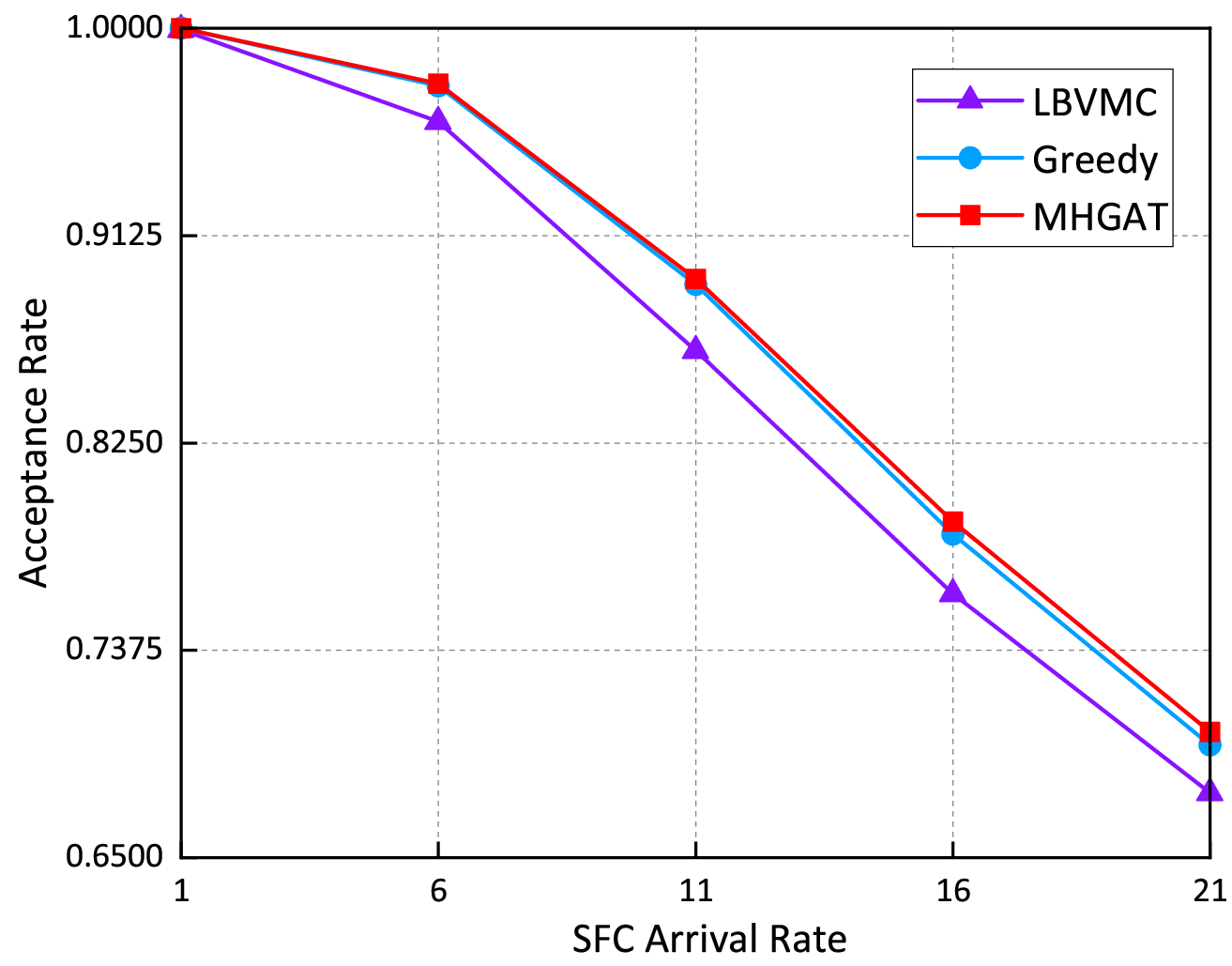}
	}
	\hfil
	\subfloat[\scriptsize{Overload ratio with different SFC arrival rates.}]
	{\includegraphics[width=1.46in]{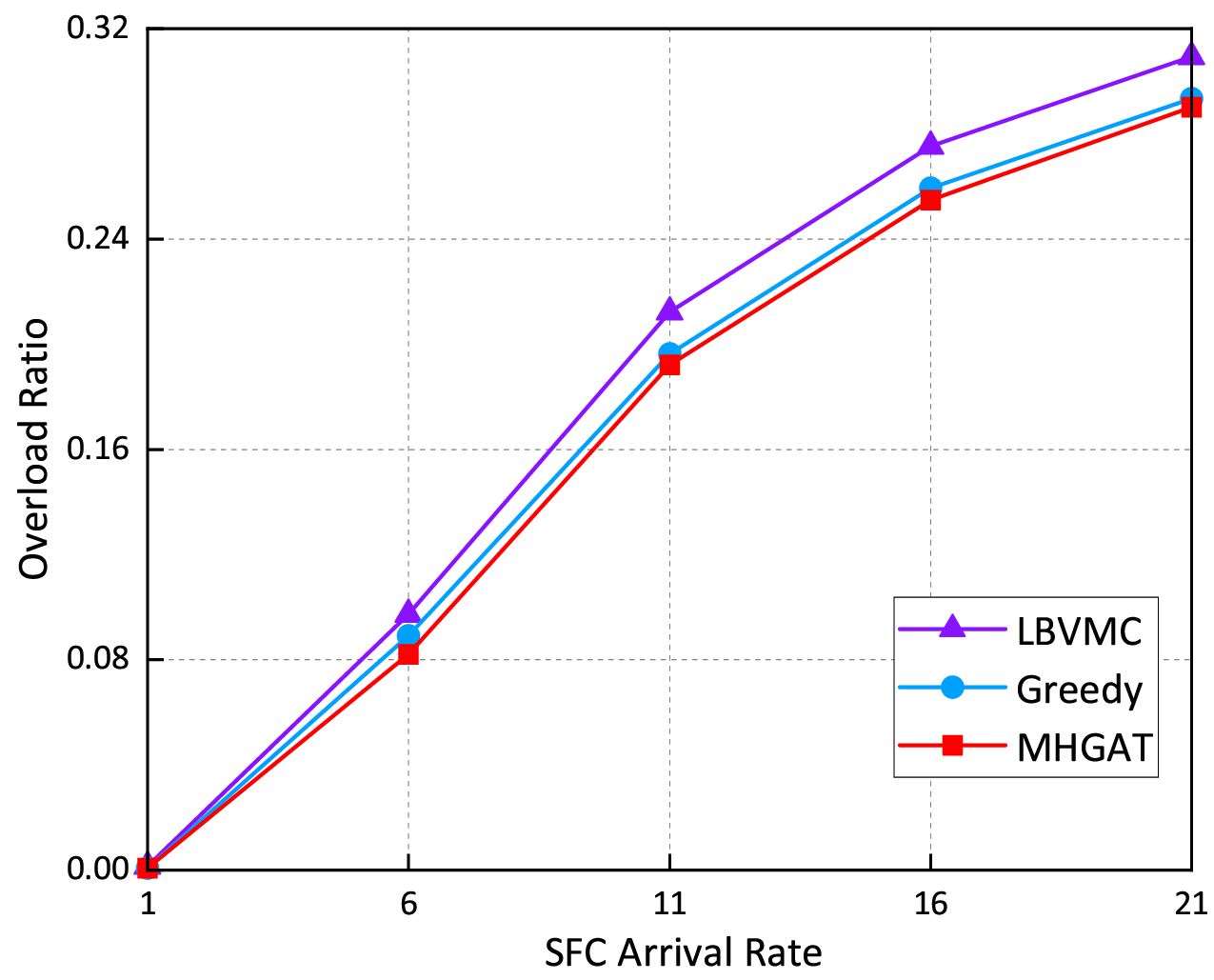}
	}
	\hfil
	\vspace{-3mm}
	\subfloat[\scriptsize{Migration loss with different SFC arrival rates.}]
	{\includegraphics[width=1.47in]{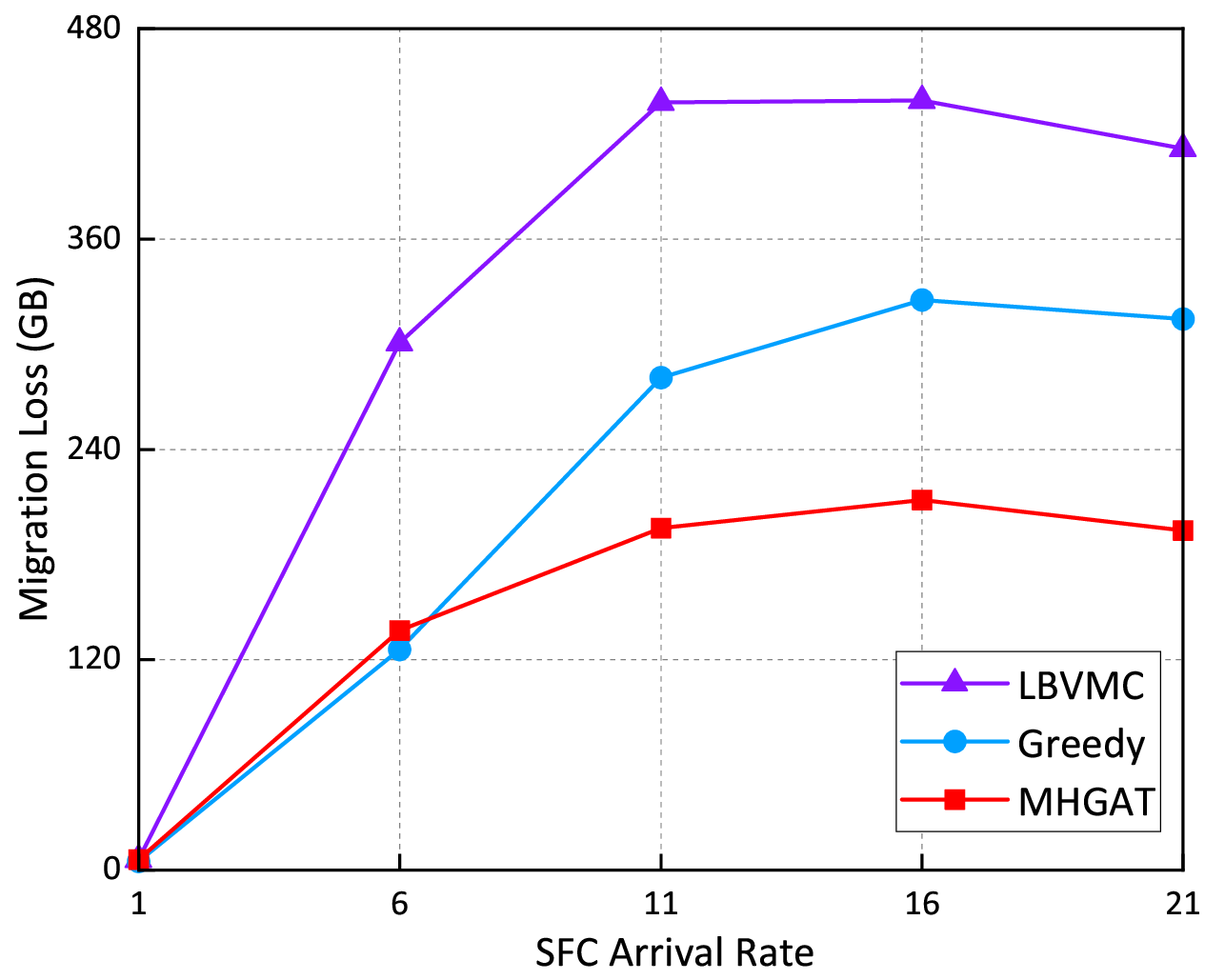}
	}
	\hfil
	\subfloat[\scriptsize{Fragmentation level with different SFC arrival rates.}]
	{\includegraphics[width=1.5in]{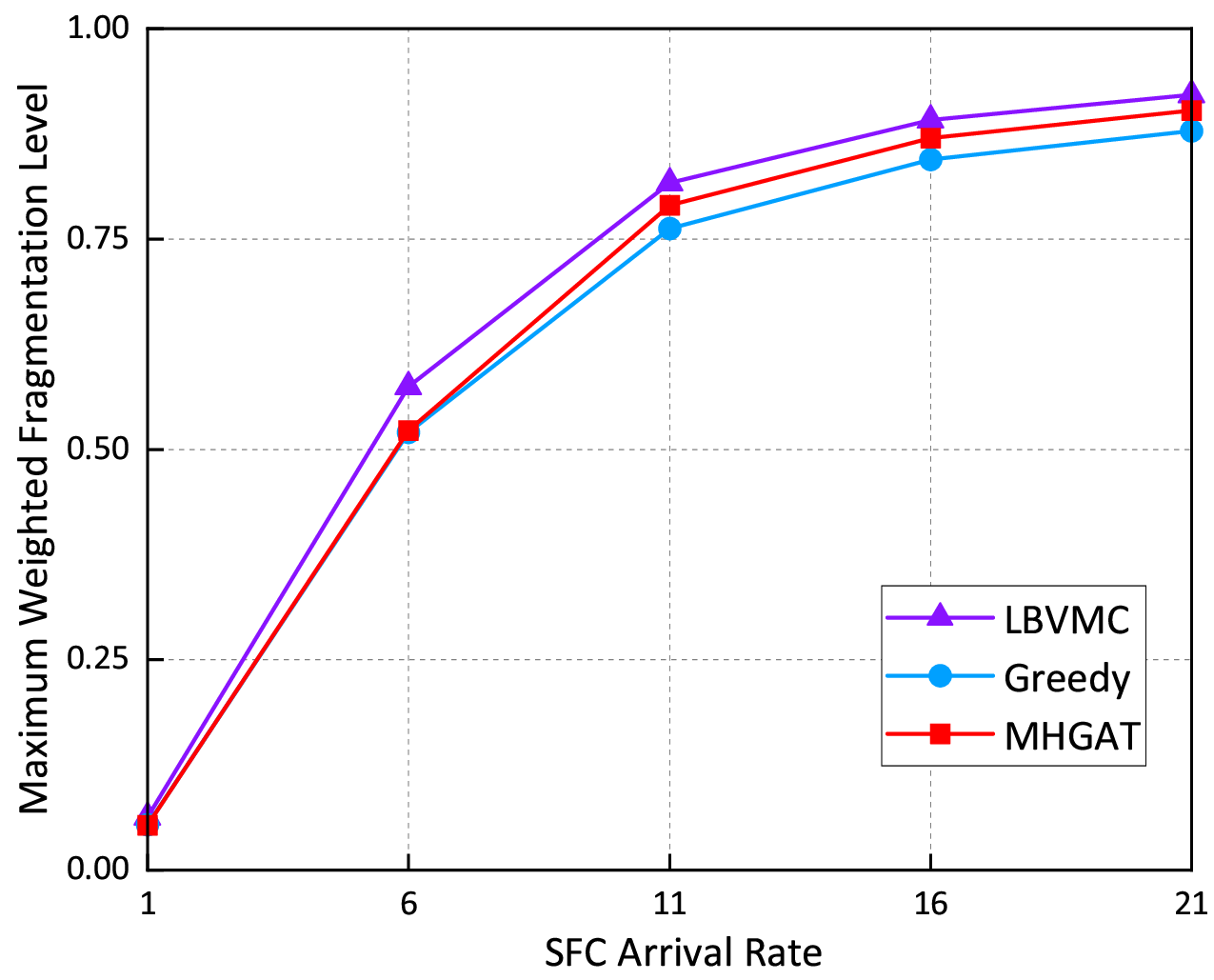}
	}
	\vspace{1mm}
	\caption{Comparison of performance metrics with different SFC arrival rates.}
	\label{fig_arr}
	\vspace{-2mm}
\end{figure*}
\begin{figure*}[!t]
	\vspace{-6mm}
	\centering
	\subfloat[\scriptsize{Acceptance ratio with different CPU capacities of nodes.}]
	{
		\includegraphics[width=1.47in]{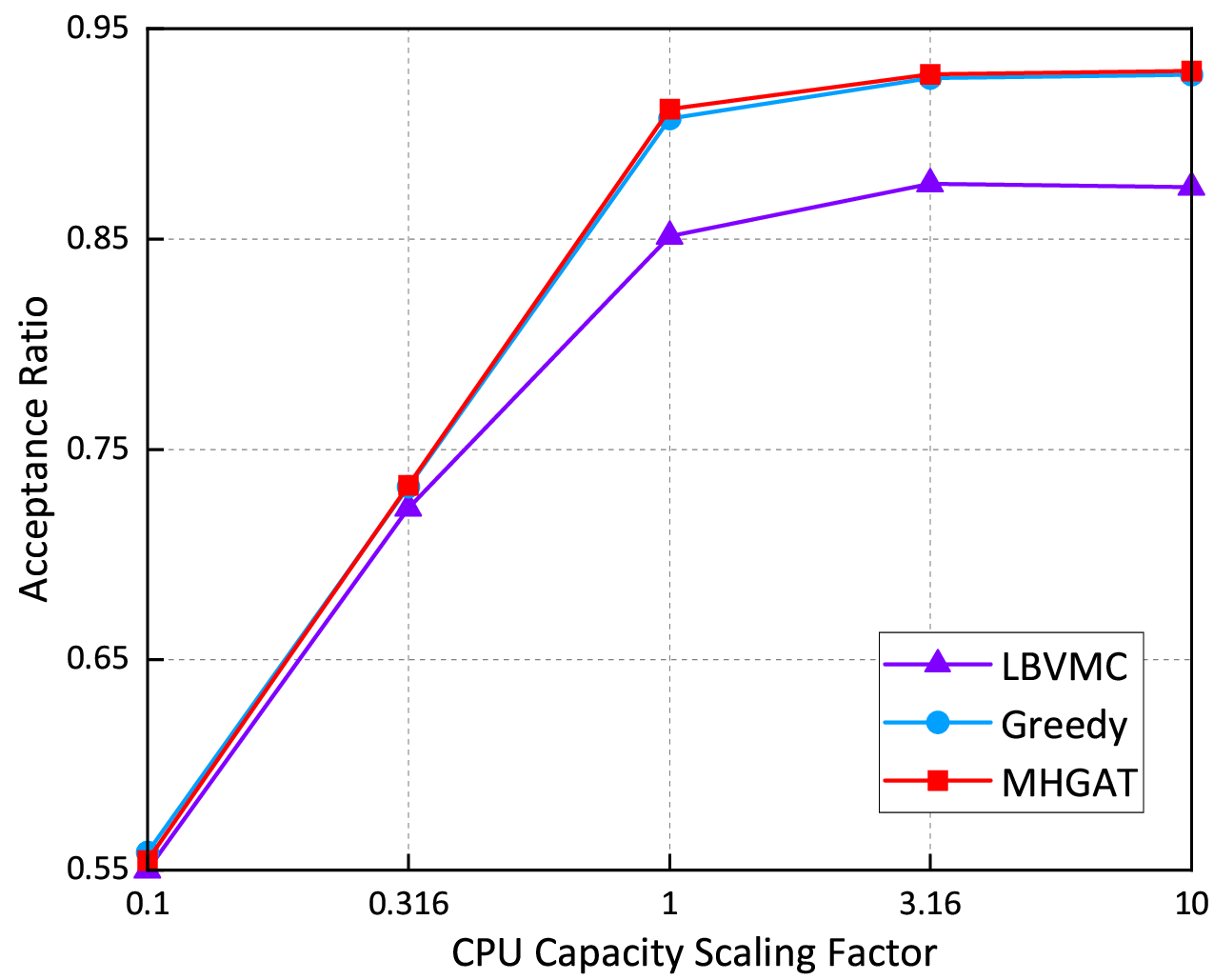}
	}
	\hfil
	\subfloat[\scriptsize{Overload ratio with different CPU capacities of nodes.}]
	{
		\includegraphics[width=1.49in]{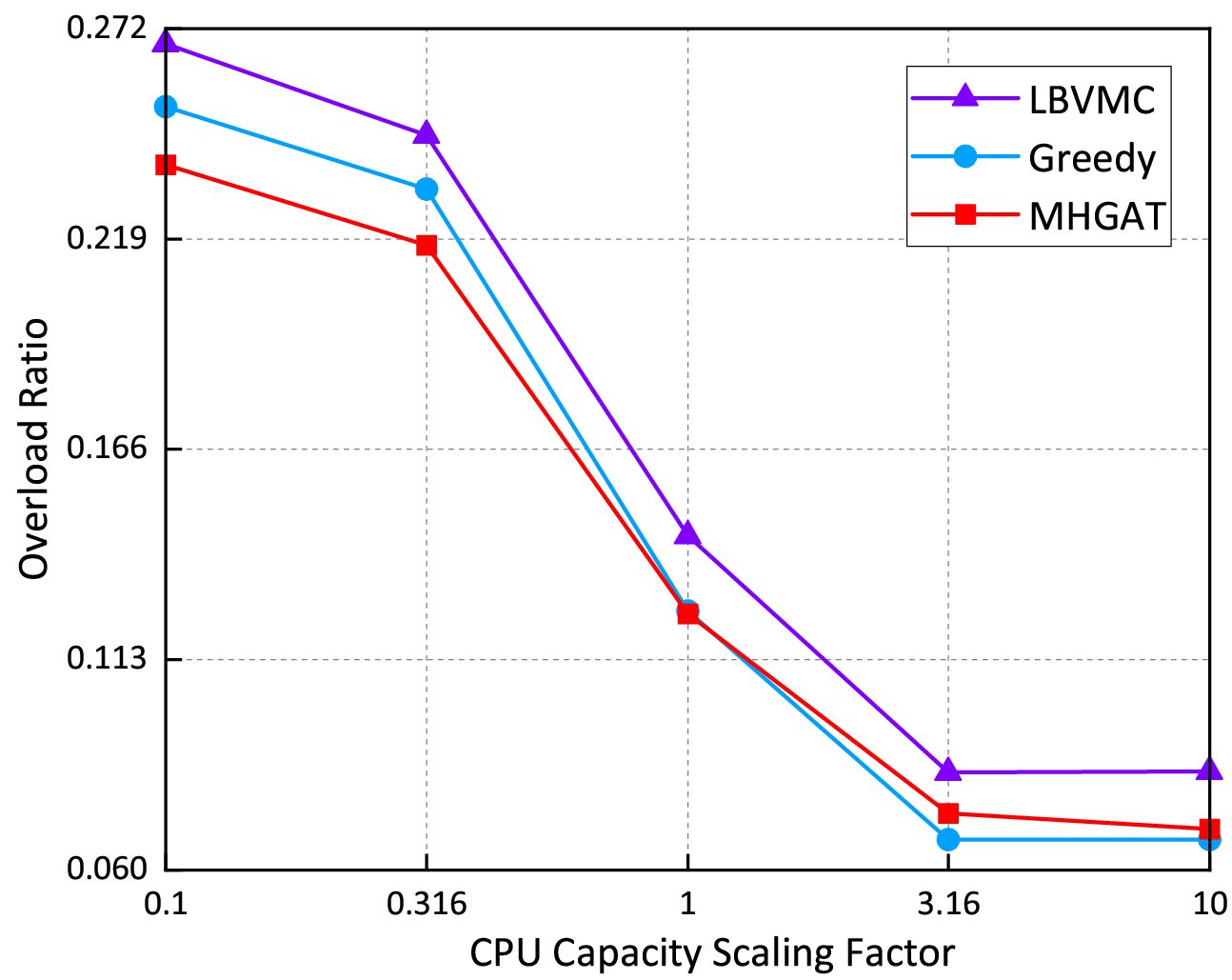}
	}
	\hfil
	\subfloat[\scriptsize{Migration loss with different CPU capacities of nodes.}]
	{
		\includegraphics[width=1.47in]{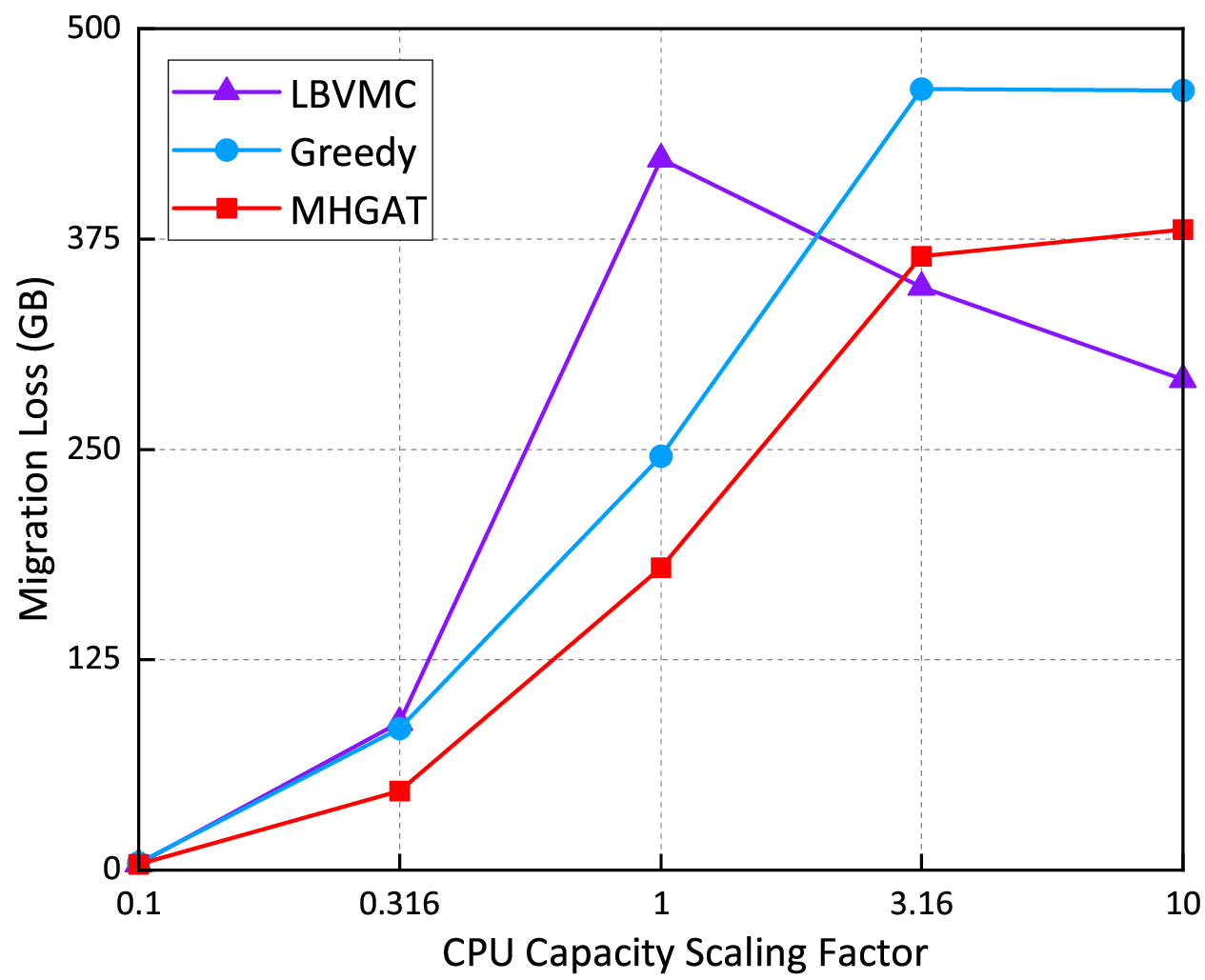}
	}
	\hfil
	\subfloat[\scriptsize{Fragmentation level with different CPU capacities of nodes.}]
	{
		\includegraphics[width=1.49in]{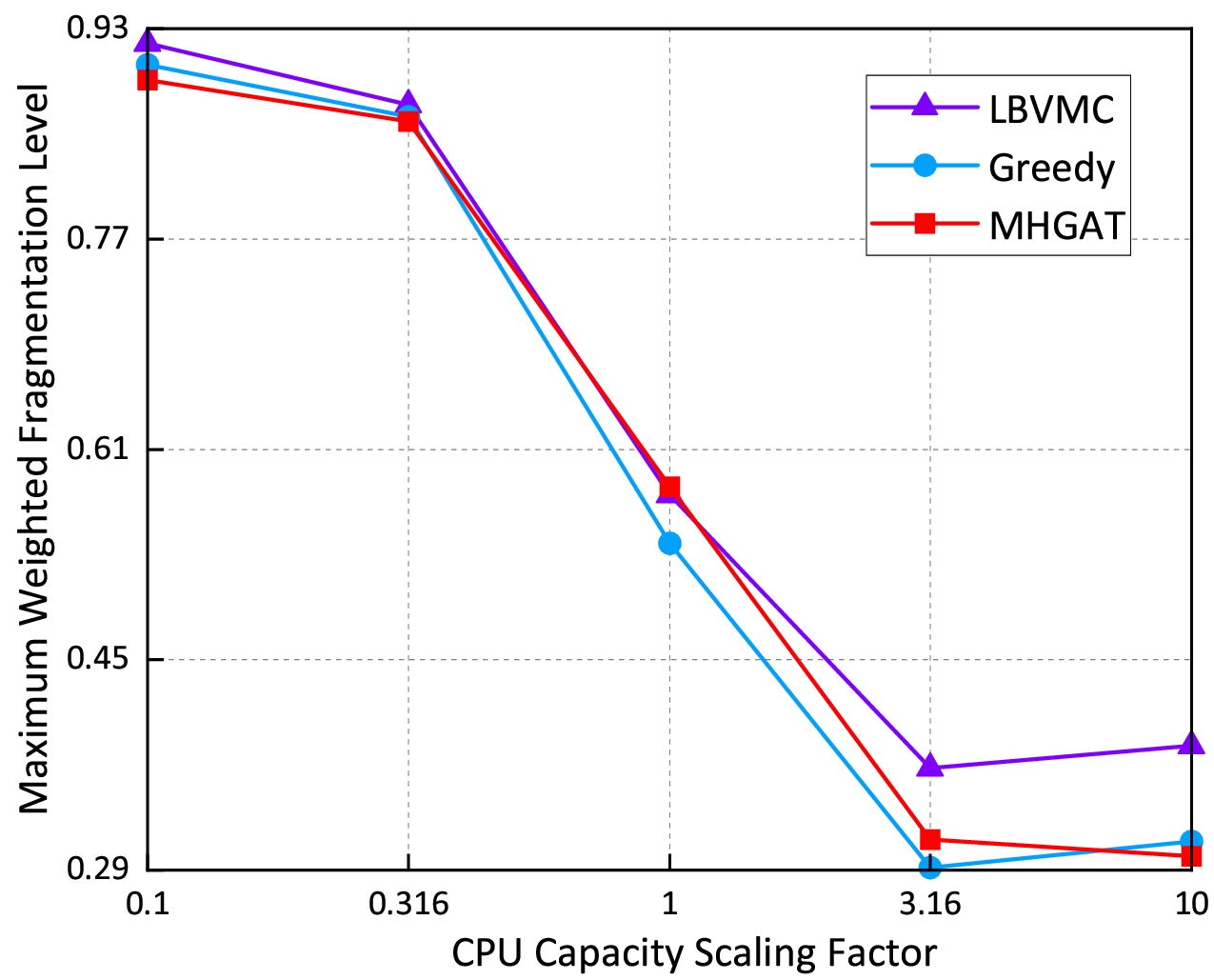}
	}
\vspace{-2mm}
	\caption{Comparison of performance metrics with different CPU capacities of nodes.}
	\label{fig_cpu}
\end{figure*}
\begin{figure*}
	\vspace{-7mm}
	\centering
	\subfloat[\scriptsize{Accpetance ratio with different memory capacities of nodes.}]
	{\includegraphics[width=1.47in]{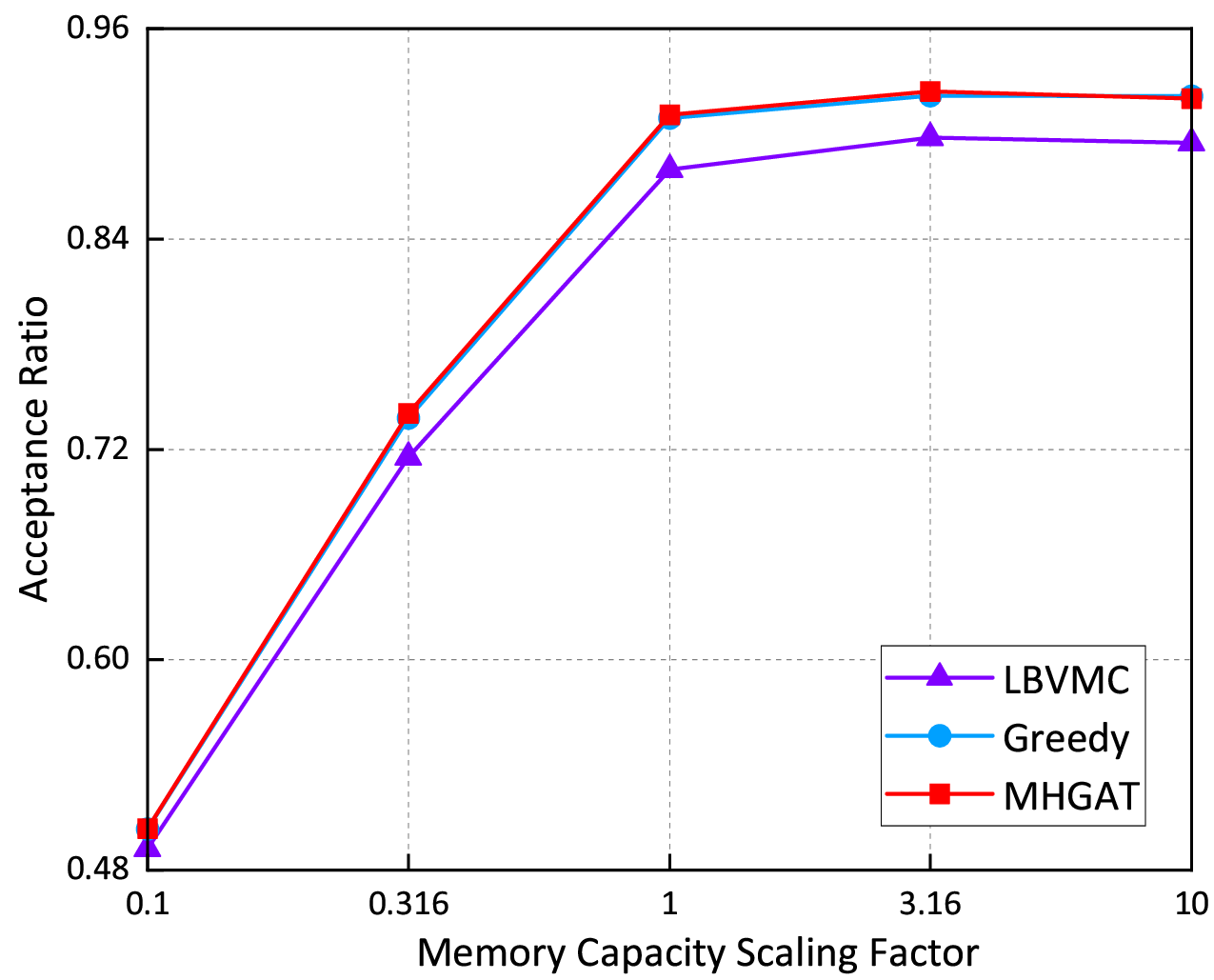}
	}
	\hfil
	\subfloat[\scriptsize{Overload ratio with different memory capacities of nodes.}]
	{\includegraphics[width=1.495in]{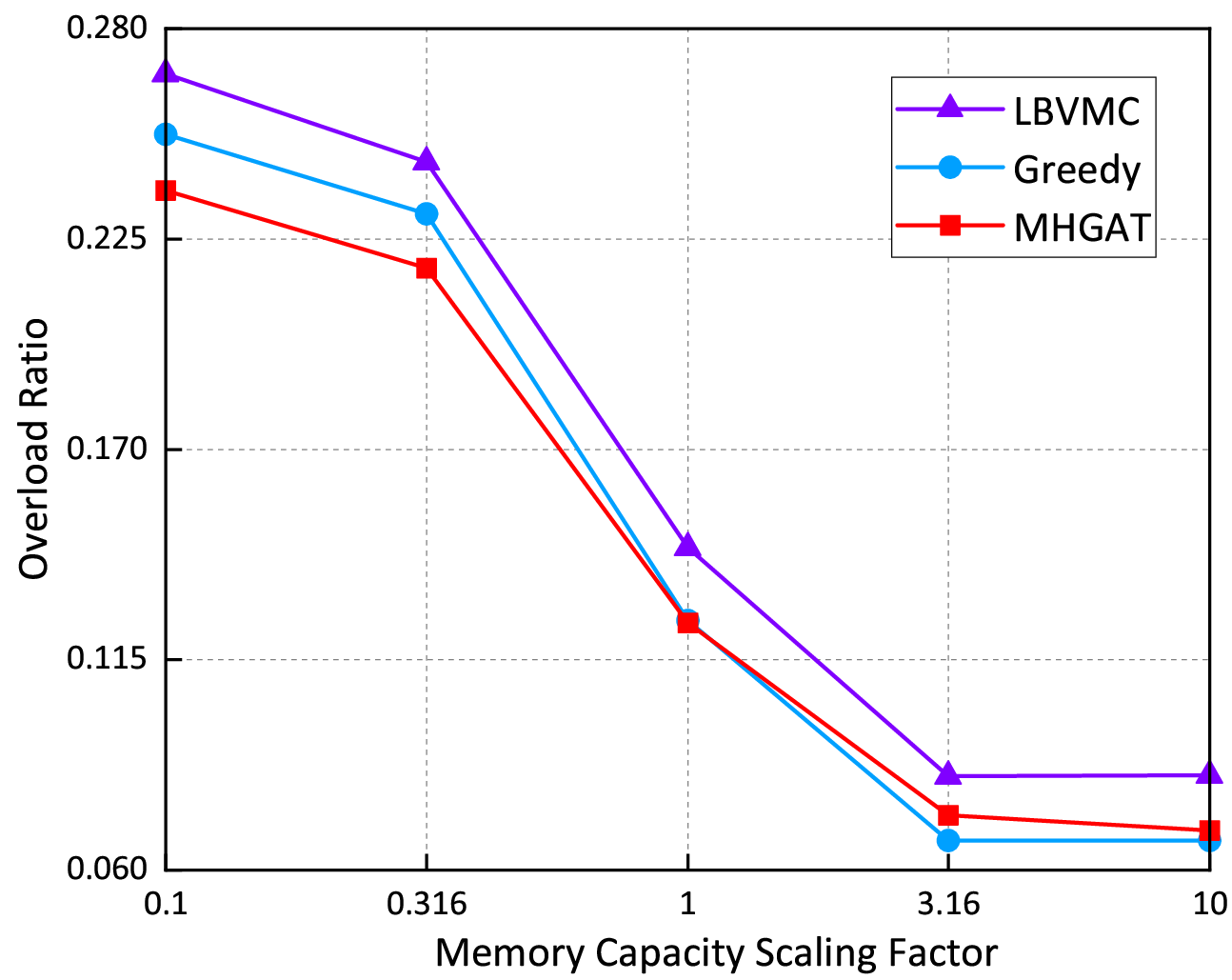}
	}
	\hfil
	\subfloat[\scriptsize{Migration loss with different memory capacities of nodes.}]
	{\includegraphics[width=1.47in]{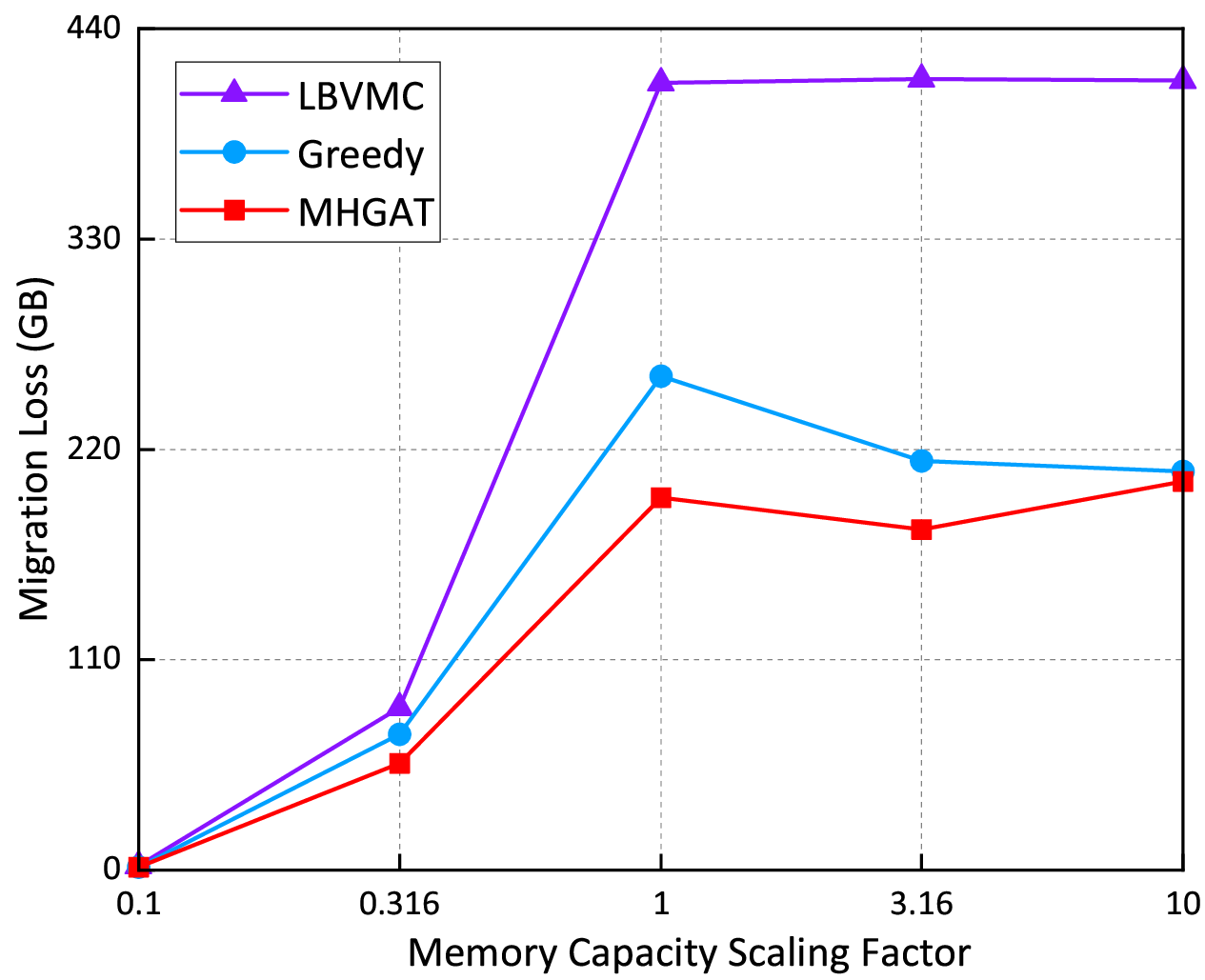}
	}
	\hfil
	\subfloat[\scriptsize{Fragmentation level with different memory capacities of nodes.}]
	{\includegraphics[width=1.5in]{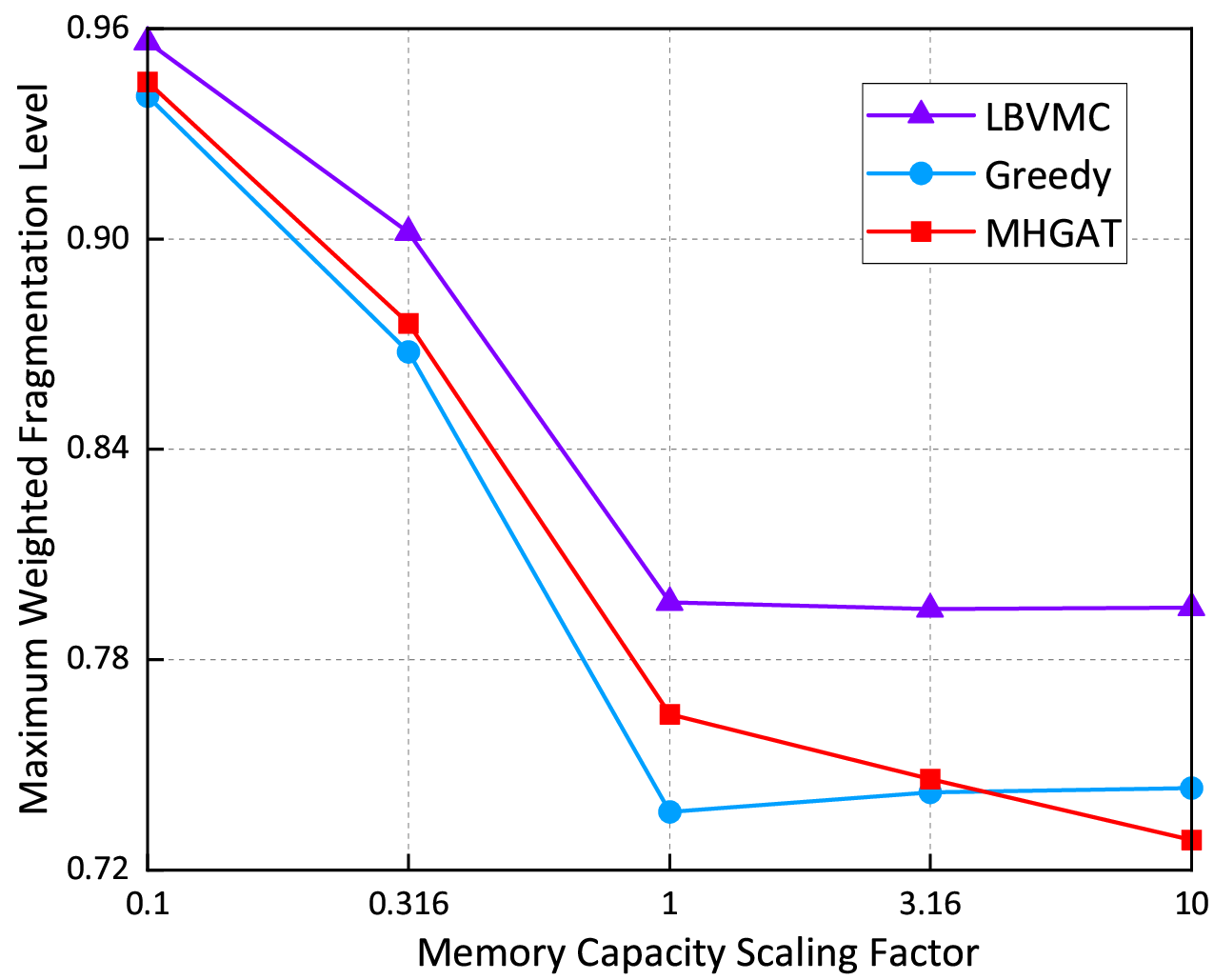}
	}
	\vspace{-3.5mm}
	\caption{Comparison of performance metrics with different memory capacities of nodes.}
	\label{fig_mem}
\end{figure*}
\begin{figure*}
	\vspace{-7mm}
	\centering
	\subfloat[\scriptsize{Acceptance ratio with different bandwidth capacities of links.}]
	{\includegraphics[width=1.465in]{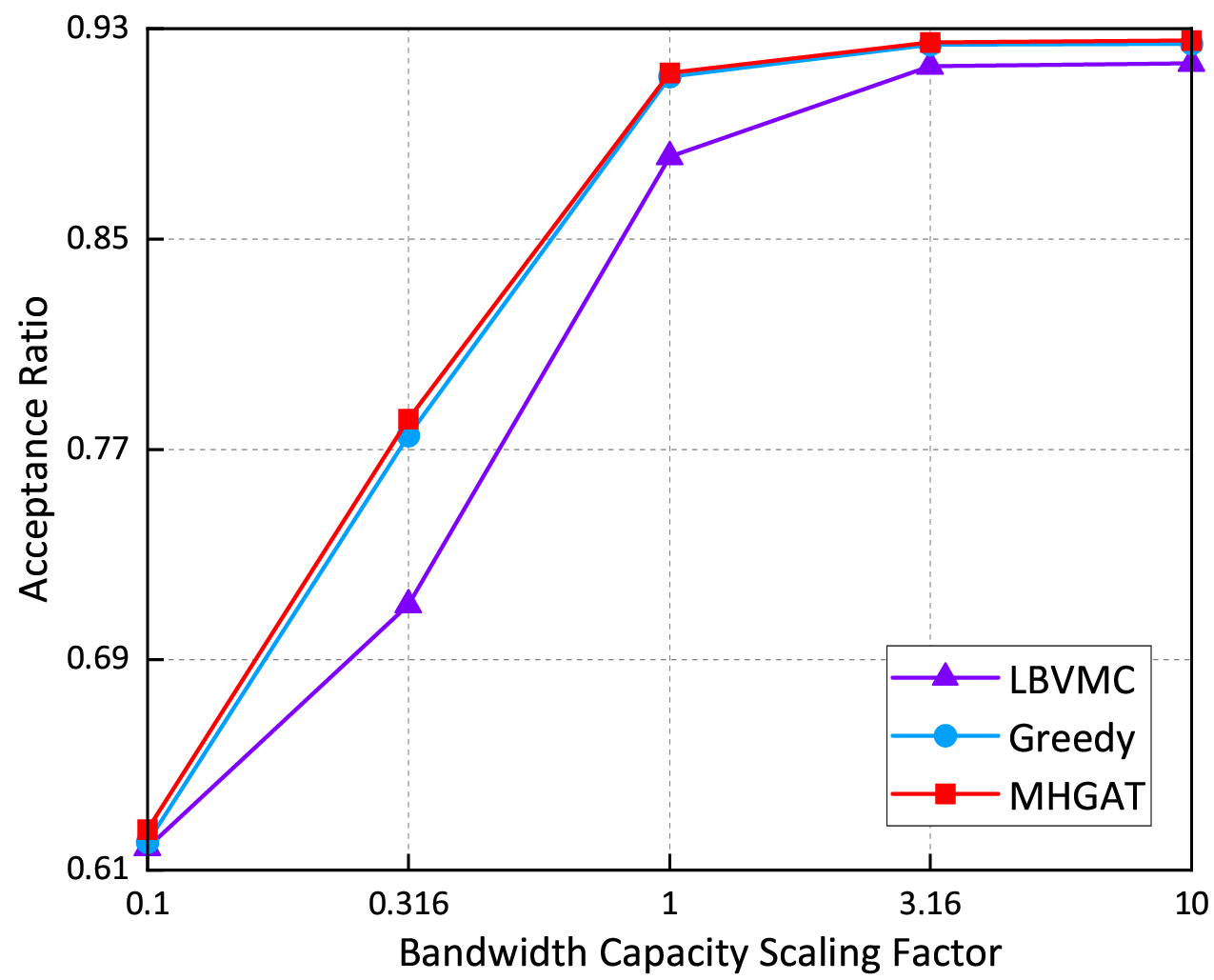}
	}
	\hfil
	\subfloat[\scriptsize{Overload ratio with different bandwidth capacities of links.}]
	{\includegraphics[width=1.475in]{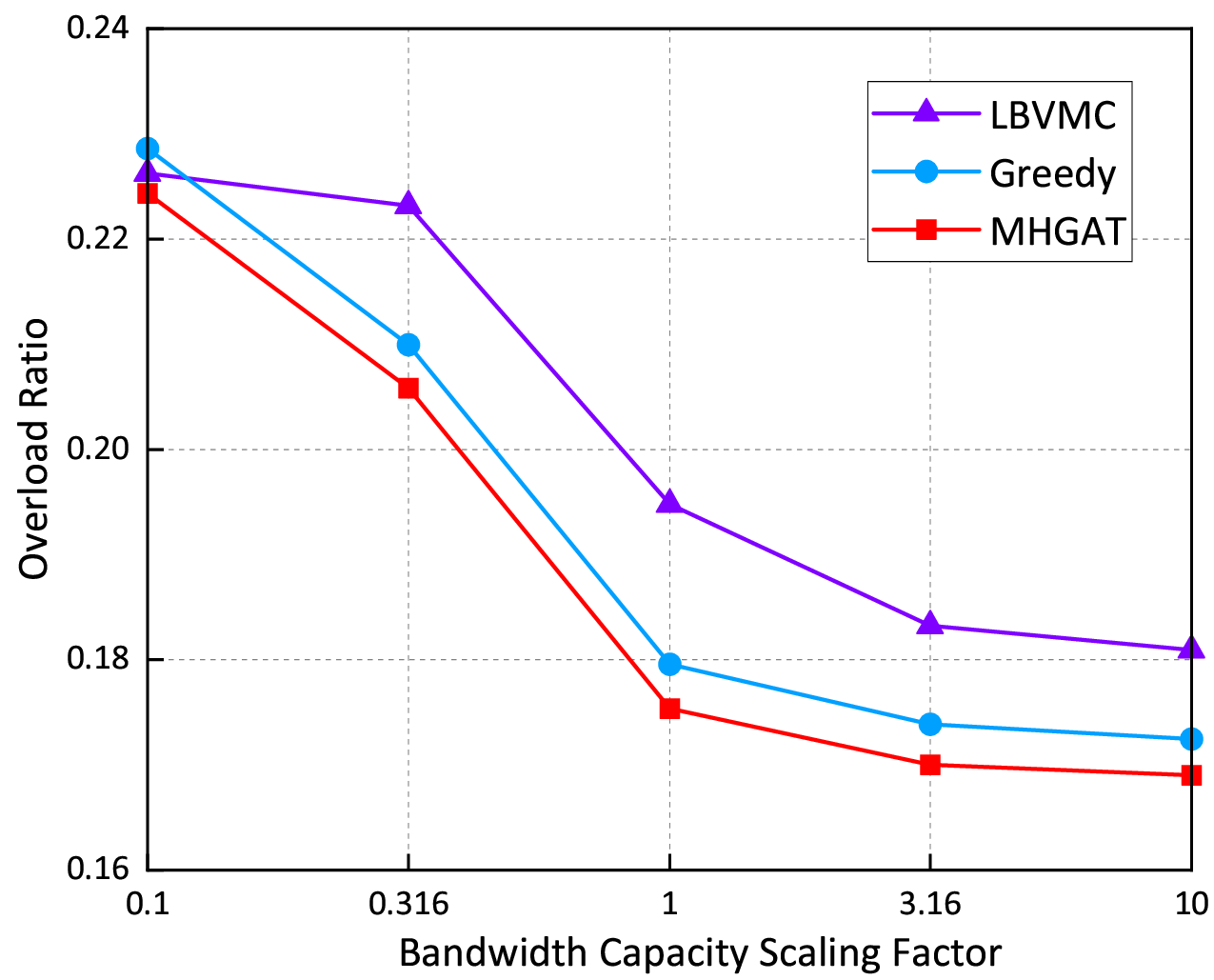}
	}
	\hfil
	\subfloat[\scriptsize{Migration loss with different bandwidth capacities of links.}]
	{\includegraphics[width=1.47in]{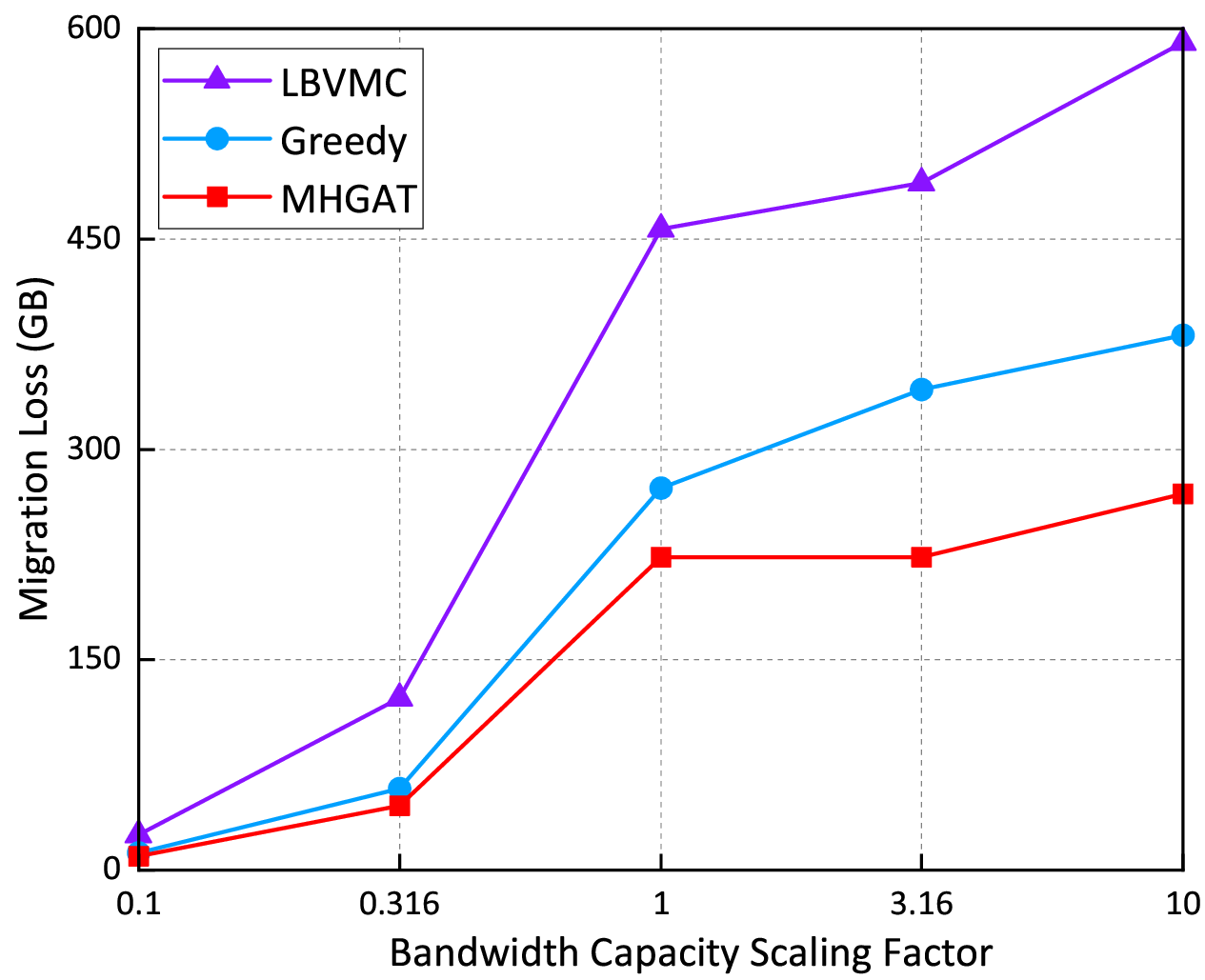}
	}
	\hfil
	\subfloat[\scriptsize{Fragmentation level with different bandwidth capacities of links.}]
	{\includegraphics[width=1.5in]{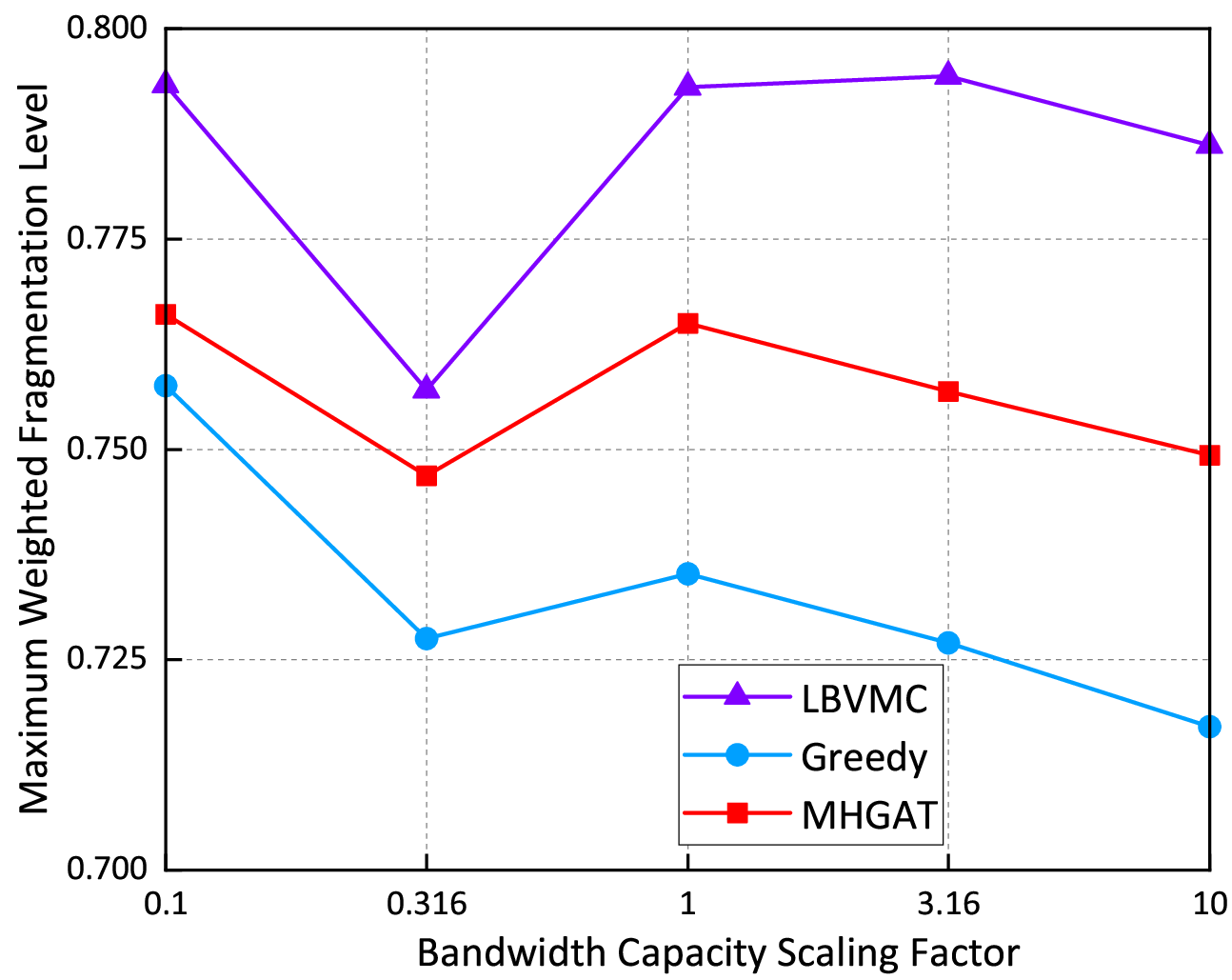}
	}
	\vspace{-3.5mm}
	\caption{Comparison of performance metrics with different bandwidth capacities of links.}
		\vspace{-6mm}
	\label{fig_bw}
\end{figure*}
\vspace{-3mm}
\subsection{Validation of MHGAT Method}
\label{sim_method}

In this subsection, the performance of the MHGAT method is validated with different system conditions. We first vary the arrival rate of SFCs to verify its performance under different pressures. Then we scale the capacity of resources in different dimensions separately to verify the ability of the MHGAT method to cope with extreme multidimensional resource pressures. Finally, we compare the running time of different algorithms.

Fig. \ref{fig_arr} illustrates the comparison of various performance metrics for different SFC arrival rates. As shown in Fig. \ref{fig_arr}(a), taking the worst algorithm performance as a benchmark, the acceptance ratio of the MHGAT method improves by an average of 12.8\% for different SFC arrival rates, which stems from the reduced fragmentation level it brings by targeting the fragmentation phenomenon. In Fig. \ref{fig_arr}(b), we can see that taking the highest overload ratio as a benchmark, the overload ratio of the MHGAT method is reduced by 30.6\% on average for different SFC arrival rates. This is attributed to the optimization of fragmentation levels by the MHGAT method, which facilitates a more balanced distribution of multidimensional resource pressures. As shown in Fig. \ref{fig_arr}(c), the MHGAT method reduces the migration loss by an average of 43.3\% for different SFC arrival rates compared to the benchmark algorithm. In addition to the reduction of migration loss by mitigating fragmentation, the passive migration triggering mode is also involved in the reduction of migration loss, while prediction-based active migration mode (e.g., LBVMC mode) generates more migration loss. Finally, Fig. \ref{fig_arr}(d) shows that the maximum weighted fragmentation level of the MHGAT method is not the lowest. This is not because our method does not work, but because the increased acceptance rate brings more SFCs, which are additional load pressures for the network. Obviously, VNF migration can only reduce the local load pressure but not the global load pressure, so the maximum weighted fragmentation level of the MHGAT method is not the lowest. Nevertheless, the targeting of fragmentation phenomena by the MHGAT method makes the network conditions less prone to require frequent migrations due to resource fragmentation, which results in fewer migration loss. 

Fig. \ref{fig_cpu}, \ref{fig_mem}, and \ref{fig_bw} show the performance of each algorithm after we scale the node CPU resources, memory resources and link bandwidth resources, respectively. It can be seen that the trend of the performance metrics of each algorithm is similar when the capacity of different types of resources is changed, but there are subtle differences. The reason for the similar acceptance rates is that it is difficult to further increase the acceptance ratio when the system load is saturated. As for the overloading ratio, a closer look reveals that the overloading rate of the MHGAT method performs better when the resource is scarce, while it performs similarly to the greedy algorithm when the resource is sufficient. This is because the resource has a higher weight in the MHGAT method when it is scarce, while its weight is diluted by the weights of other resources when it is sufficient. In migration loss comparisons, the MHGAT method stably outperforms the other algorithms in most cases due to the fact that defragmentation makes the node and link resources slower to generate resource fragmentation. Finally, we compare the fragmentation levels of the different methods, and the MHGAT method does not achieve a clear advantage since this metric is affected by real-time fluctuating resource requirements and different SFC acceptance scenarios.

\begin{figure}[!t]
	\centering
	\includegraphics[width=3in]{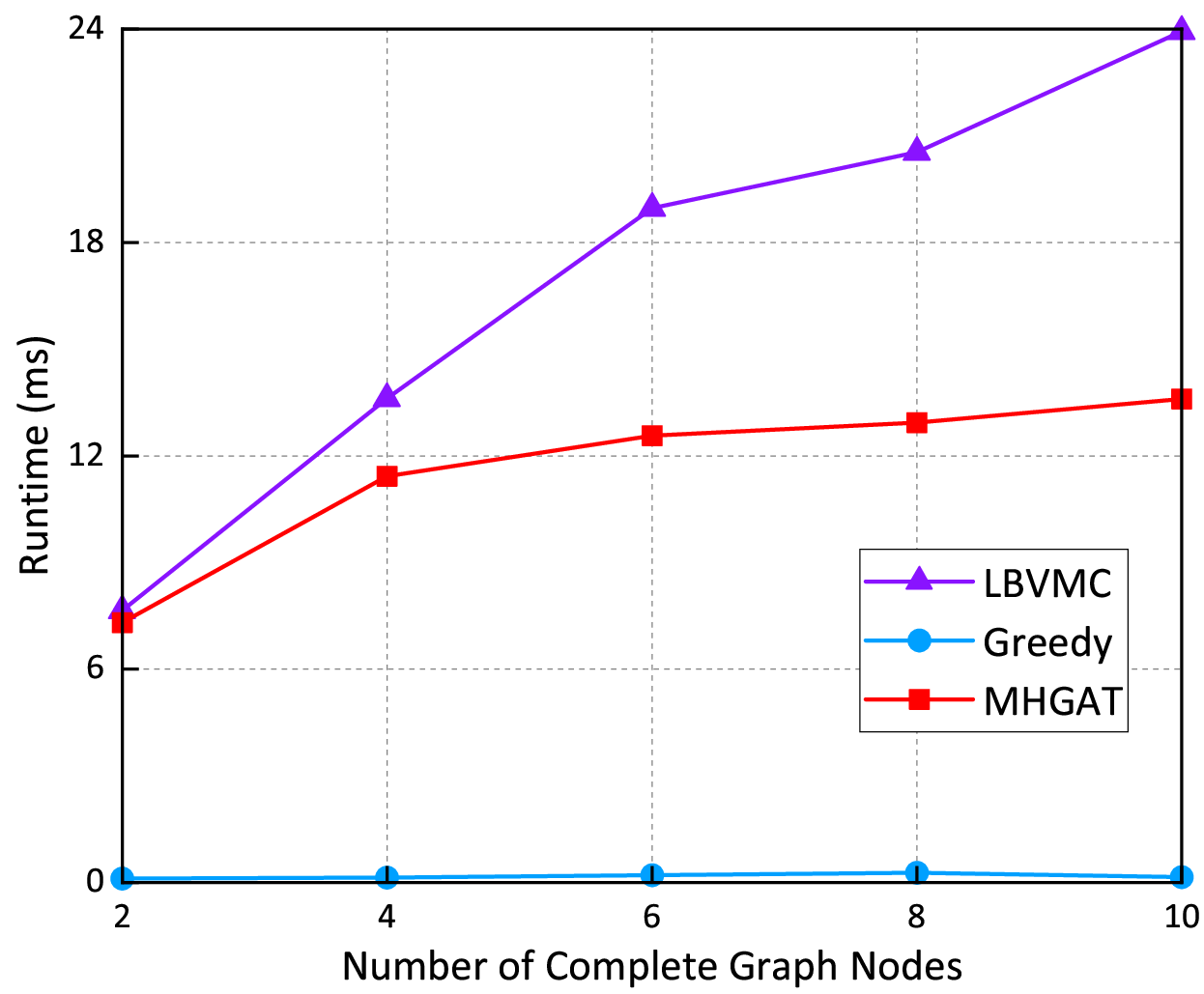}
	\caption{Comparison of runtime in complete graph topologies with different numbers of nodes.}
	\label{fig_runtime}
	\vspace{0mm}
\end{figure}

In our last simulation, we tested the runtime of each algorithm in the full graph topologies with different numbers of nodes, where a complete graph refers to a graph in which every pair of nodes is connected by an edge. From Fig. \ref{fig_runtime}, it can be seen that the MHGAT method has a slower increase in runtime compared to LBVMC and can produce results in 14 \textit{ms} in a complete graph topology with 10 nodes and 45 links. This is due to the fact that the number of hidden layer parameters of the neural network is fixed and the complexity comes only from the input and output layers. Nevertheless, the greedy algorithm still shows a time efficiency that is hard to beat.
\vspace{-3mm}
\section{Conclusion And Outlook}
In this paper, we propose a new metric to quantify network resource fragmentation and a novel neural network model MHGAT to capture network resource features. Based on the metric and migration loss, we model the VNF migration problem as an integer nonlinear programming problem and propose a deep learning method to solve it based on the MHGAT model. By considering the multi-hop neighbor characteristics of the nodes, the proposed metric is able to extract network resource fragmentation information effectively. In addition, the proposed MHGAT model can efficiently extract a large range of network resource features through neural network design for multi-hop information and provide an efficient strategy for VNF migration. The simulation results verify the effectiveness of the proposed metric and neural network model, and confirm that the MHGAT method improves the service acceptance rate by an average of 12.8\%, reduces the overload rate by an average of 30.6\% and the migration loss by an average of 43.3\% for different SFC arrival rates compared to benchmark algorithms.

For future work, we plan to further investigate the issue of network resource fragmentation in MEC. Due to potential issues such as user mobility and dynamic topology at the network edge, VNF migration strategies need to further consider the impact of these factors to minimize migration loss and optimize resource allocation. In addition, adjusting network resource allocation through VNF scaling is also one of our goals.
\bibliographystyle{ieeetr}
\bibliography{ref.bib}
\end{document}